\documentclass[letterpaper,twocolumn,11pt]{article}

\newlength\titlebox
\usepackage{times}
\usepackage{helvet}
\usepackage{courier}
\usepackage{url}
\usepackage{graphicx}
\usepackage{color}
\usepackage{xcolor}    
\usepackage{multirow,rotating}
\usepackage{soul}
\usepackage{booktabs}
\usepackage{comment} 
\usepackage[]{authblk}  

\usepackage[font=small]{caption}
\usepackage{abstract}
\date{July 8, 2013}
\usepackage[compact]{titlesec}
\title{Rising tides or rising stars?: Dynamics of shared attention on Twitter during media events}

\author[1,*]{Yu-Ru Lin}
\author[1]{Brian Keegan}
\author[2]{Drew~Margolin}
\author[1,2]{David Lazer}
\affil[1]{College of Social Sciences and Humanities, Northeastern University, Boston, MA 02115, USA}  
\affil[2]{College of Computer and Information Science, Northeastern University, Boston, MA 02115, USA}
\affil[*]{To whom correspondence should be addressed. Email: \protect\url{yuruliny@gmail.com}}

\begin{document}

\newcommand{\specialcell}[2][l]{%
  \begin{tabular}[#1]{@{}l@{}}#2\end{tabular}}
\newcommand{\vrtLbl}[2]{\multirow{#2}{1ex}{\begin{sideways}#1\end{sideways}}} 

\def\figfilepath{figs}
\graphicspath{{\figfilepath/}}

\def\todo#1{{\color{red}\textbf{}}}
\def\yrl#1{{\small\color{orange}\textbf{}}}
\def\bk#1{{\small\color{purple}\textbf{}}}

\twocolumn[
    \maketitle
    \vspace{-3em}
    \begin{onecolabstract}
        \vspace{-1em}
``Media events'' such as political debates generate conditions of shared attention as many users simultaneously tune in with the dual screens of broadcast and social media to view and participate. Are collective patterns of user behavior under conditions of shared attention distinct from other ``bursts'' of activity like breaking news events? Using data from a population of approximately 200,000 politically-active Twitter users, we compare features of their behavior during eight major events during the 2012 U.S. presidential election to examine (1) the impact of ``media events'' have on patterns of social media use compared to ``typical'' time and (2) whether changes during media events are attributable to changes in behavior across the entire population or an artifact of changes in elite users' behavior. Our findings suggest that while this population became more active during media events, this additional activity reflects concentrated attention to a handful of users, hashtags, and tweets. Our work is the first study on distinguishing patterns of large-scale social behavior under condition of uncertainty and shared attention, suggesting new ways of mining information from social media to support collective sensemaking following major events.
    \end{onecolabstract}
]

\section{Introduction}\label{sec:intro}
Twitter plays an increasingly important role disseminating information for both breaking news events as well as ceremonial media events. Examples of these media events include political campaigns, sporting events, and television shows where the dates are scheduled and the stakes are known. Increasingly, these media events invite viewers to ``dual screen'' by using their mobile devices to share their reactions on social media during live television broadcasts. These back-channel forms of communication facilitate processes of information sharing as audiences respond to narratives driven by the content on both screens.

Social media have the potential to allow individuals to openly communicate with large audiences over topics of shared interest. However, socio-technical systems like Twitter are also characterized by large disparities where the vast majority of activity is concentrated around a few ``stars.'' This would suggest media events should reproduce the concentrated attention towards elite users rather than bringing in new voices. However, the aphorism ``a rising tide lifts all boats'' reflects how changing systems can benefit all the members of that system.

This paper addresses two related questions. The first examines whether shared attention changes user behavior: do Twitter users behave differently during a media event as compared to breaking news events or ``normal'' time? On one hand, media events could simply reproduce existing behavioral patterns at larger volumes of activity but on the other hand, they could also cause significant changes in users' communication patterns of using hashtags, mentions, and retweets. If these behavioral differences exist, this leads to a second question: are behavioral differences the result of behavior shifting across the entire population or do some users preferentially benefit from behavioral changes during media events? Shared attention to an event could let new voices emerge and create a ``rising tide'' where behavior across all users changes in similar ways. Alternatively, shared attention to an event could reinforce the popularity of elite users and create ``rising stars'' that drown out alternative voices. 

Drawing on the concepts of ``media events'' and ``social foci'' from media studies and sociology, we propose a theory of \textit{media event-driven behavioral change} to explain how conditions of shared attention alter collective behavior in social media. Our research design uses eight events related to U.S. politics that occurred between late August and mid-October 2012 as natural experiments to analyze changes in the behavior of approximately 200,000 politically-engaged Twitter users. Media events like the national political conventions and presidential debates are compared to the news events surrounding the Benghazi attacks and ``47\%'' controversy as well as four ``'typical'' days preceding each of the debates. Examining the changes in the rate and concentration of communication patterns, connectivity, and user responsiveness, the findings provide support for our theory of media event-driven behavioral change by demonstrating that (1) fundamental patterns of social media activity are significantly altered during media events compared to non-media events and (2) these differences are largely attributable to an increasing concentration of attention to elite users rather than an increasing distribution of activity and attention across all users.

\section{Background}\label{sec:background}
``Media events'' are defined by Dayan and Katz as ceremonies or performances that interrupt routines and invite audiences to join in a collective experience~\cite{dayan1992media}. These events monopolize media coverage and displace other regularly scheduled forms of media, are transmitted in real time as they occur, are organized by actors outside of the control of the media, are pre-planned and advertised to viewers in advance of their occurrence, are treated by the media with reverence and ceremony, and enthrall large audiences through a norm of shared viewing. Examples of media events include sporting events and political speeches that temporarily create an occasion for large portions of society to ignore divisions of geography, culture, and status and focus together on a single event that displaces all other events. 

Other events share some but not all of these features. News events may be large enough to displace coverage, occur outside of media control, and are covered live, but they are not pre-planned and thus lack the ceremonial aspects or large, enthralled audiences. Examples of news events include disasters and scandals where information is scarce and the implications are uncertain. Nevertheless, significant changes in social media use are observed during major events such as professional sports~\cite{nichols2012summarizing}, television shows~\cite{ciulla2012beating}, and natural disasters~\cite{vieweg2010microblogging}.  These back-channel sources of communication during other events highlight the potential of these technologies to support rapid information sharing, broad participation, and collective sensemaking as audiences develop and respond to narratives which are driven by the content they are consuming on both their television and mobile device screens~\cite{shamma2009tweet}. 

The extensive adoption of ``dual screening'' practices also influences how citizens are consuming political information: 11\% of live debate watchers ``dual screened'' the debate by using social media on a computer or mobile device simultaneously with watching television coverage~\cite{pew2012dualscreening}. Users' tweets during political debates encode important shifts in sentiment~\cite{diakopoulos2010characterizing} and political candidates' behavior on Twitter predicts their likelihood of winning elections~\cite{livne2011party}. The spread of political hashtags is also a type of complex contagion~\cite{romero2011differences}.

Prior work has examined how temporal patterns of collective attention and information propagation on social media are affected by users' finite attention influence when attending to competing information sources~\cite{weng2012competition}, the novelty of the information itself~\cite{wu2007novelty}, the reactions of early users~\cite{lerman2010using}, and the balance between rates of spreading versus forgetting~\cite{crane2008robust}. Retweets play an important role both labeling content and signaling membership in a community~\cite{yang2012we}. Users are more likely to retweet content that requires less cognitive effort to act upon by focusing their attention on the most recent and visible content~\cite{counts2011taking,hodas2012visibility}. 

With some exceptions~\cite{lehmann2012dynamical}, analyses of collective attention and information spreading have generally assumed the system is operating at a steady state and do not account for exogenous shocks that change many users' communication and information sharing behaviors. Does dual screening during media events lead to significant changes in behavioral patterns across a population of politically-active Twitter users? If it does, how do these differences manifest themselves across different levels of this population? 


\section{Media event-driven behavioral change}\label{subsec:differences}
Media events could potentially create larger volumes of activity without altering the type or distribution of underlying behavioral patterns. For example, while there may be more people generating more tweets, the underlying rates of mention, hashtag, and retweet use could remain unaltered. However, we argue the social and cultural contexts of information consumption during media events interact with repertoires of technology use such as dual screening and lead to what we call \textit{media event-driven behavioral change}. 

The conditions of shared attention created by media events displace normal content and become foci around which individuals collectively orient their behavior and organize new social relations~\cite{feld1981focused}. When there are no media events or conditions for shared attention, individuals' contexts and activities are fragmented among a variety of different places, groups, and interests. Their aggregate patterns of using language, hashtags, mentions, and retweets exhibit some behavioral regularities~\cite{kwak2010twitter,golder2011diurnal}, but are largely uncorrelated between different foci and the communities surrounding them.

Media events are not simply live events with large audiences, they are rare and enthralling and ceremonial occasions that demand shared attention. They disrupt prevailing fragmented behavioral patterns by creating a single focal activity to which people pay attention. This shared attention in turn displaces normal contexts and the behaviors associated with them. In effect, the media event is an external force that causes the population of users to go through a phase transition from a relatively disordered state to a highly ordered state~\cite{castellano2009statistical}. This ``highly ordered'' state is characterized by users attending to a single shared focal activity rather than many focal activities and behaving in unusually active ways such as dual screening. Furthermore, users in this highly ordered state produce and consume information from an audience that is reflexively aware of their shared exposure to the same information. In other words, everyone knows that everyone is watching the same thing.

Like television in its early years~\cite{lang1953unique}, social media can magnify the significance viewers ascribe to media events and create a liminal space inaccessible to the audience members physically present at the event, viewers watching only television coverage of the event, or users solely monitoring social media. These liminal spaces generate novel understandings and emerging consensus on the meaning of the media event. For example, concepts like ``big bird'', ``binders'', and ``bayonets'' are meaningless unless they are interpreted in the context of their enthusiastic sharing and improvisation by social media users during media events like the 2012 U.S. Presidential debates. This single shared focus and intense activity will qualitatively alter prevailing behavioral patterns and communication structures.


We expect that the enthralling nature of media events will not only increase tweet volumes, but events characterized by high levels of shared attention will change behavioral patterns such as using retweets, mentions, and hashtags. Users participating in the liminal, dual screening space between traditional and social media will be more receptive to content from other users also engaged in dual screening. 

\subsection{Behavioral changes resists concentration}\label{subset:concentration}
If behavioral differences exist during the shared attention of media events, are these differences spread across the population or do some segments like elites preferentially benefit from this increased attention? Users' increased attention to a few dominant users could lead to rapid convergence around topics as elites guided collective attention to specific interpretations and frames. For example, users retweeting elite users would lead to increased concentration of novel retweets as individuals adopted these leaders' messages. Alternatively, users' may become more receptive to all types of messages in which case replies, mentions, and retweets will increase but will not change the concentration of mentions, replies, hashtags, or retweets.

We argue that media event-driven behavioral change will disrupt patterns of behavior across all classes of users and will prime users to share and communicate more. Conditions of shared attention will create opportunities for new and more diverse voices to be heard as fragmented social foci begin to merge and overlap: average users making average tweets who are never retweeted more than two or three times might receive dozens of retweets during a media event. Conversely, elite users who already tweet regularly will see their share of attention decline as more users have larger net changes in activity than the elites, which in turn increasingly displaces elite content. Thus, despite increasing levels of activity and changes in aggregate behavioral patterns, the merging of fragmented social foci during media events should cause the structure of mentions, retweets, and hashtags to become increasingly decentralized.

\section{Research design}\label{sec:design}
We identified eight events related to the 2012 U.S. Presidential campaign that occurred over the approximately six-week period of time between late August and mid-October. Six media events were identified during this time: the Republican National Convention (RNC) from August 27 through August 30 (``CONV 1''), the Democratic National Convention (DNC) from September 4 through 6 (``CONV 2''), three debates on October 3 (``DEB 1''), 16 (``DEB 3''), and 22 (``DEB 4'') involving the presidential candidates, and single vice presidential debate on October 11 (``DEB 2''). We contrast these media events with two news events that occurred in the same span of time: the terrorist attack on the American consulate in Benghazi that killed Ambassador J. Christopher Stevens on September 11 (``NEWS 1'') and the release on September 18 of a video in which Mitt Romney argues ``47 percent'' of Americans are ``dependent upon government'' (``NEWS 2''). Both of these news events were major stories that dominated media attention for several days. 

\begin{table*}[!htb]
    \centering
    \begin{tabular}{ l llll }
    \toprule
& PRE & NEWS & CONV & DEB \\
    \midrule
description & Pre-debate baseline & \specialcell{Benghazi attack,\\ 47\% controversy} & \specialcell{Republican Nat'l Conv. \\Democratic Nat'l Conv.} & Presidential debates \\
time & \specialcell{4 days before each debate\\ (20:00-20:00 EDT)} & 
\specialcell{2-day news cycle \\(14:00-14:00 EDT)} &
\specialcell{3 days \\(08:00-14:00 EDT)} & 
\specialcell{4 hours \\(20:00-02:00 EDT)} \\
duration & 96 hours$\times$4 & 48 hours$\times$2 & 66 hours$\times$2 &
6 hours$\times$4\\
\specialcell{peak tweet volume} & 441,168&131,636 & 296,138&1,591,513 \\
\specialcell{peak unique users} & 58,823 & 30,684& 38,864& 114,663\\
\specialcell{political relevance ratio} & 0.08 & 0.16 & 0.50 & 0.63\\
 shared attention & none & low & medium & high\\
    \bottomrule
    \end{tabular}
    \caption{ Summary of datasets.
}
    \label{tab:pop}
\end{table*}

Together, these eight events make up a continuum of varying shared attention: (1) news events that should exhibit low levels of media event-driven behavioral changes since these have diffuse audiences and low mutual awareness of audience members, (2) the national political conventions that should exhibit medium levels of media event-driven changes since partisans selectively expose themselves to the conventions reflecting their political beliefs, and finally (3) the debates that should exhibit the highest levels of media event-driven change as its live and ceremonial nature drives intense shared interest.

Each of these types of events provide conditions of a natural experiment to understand how large and exogenous shocks cause changes in a well-defined population. To provide a baseline for these events, we compared the activity of this population during the peaks of these events to the peak levels of activity four days before each of the debates when there were no major media or news events. The six media events and two news events that make up the treatment group and four ``typical'' events that make up the control group give us a total 12 distinct observations of how the behavior of the same population of Twitter users changed during these events. Although tweet volumes vary regularly throughout the week, these pre-events fell on different days of the week during each of their 96-hour windows reducing the systematic bias of these events. In general, users' behavior during the ``typical'' time preceding the debate events might have been impacted by the excitements of expected debates and other campaign events, leading to a conservative comparison of changing behavior. This conservative comparison is more appropriate because it ensures that the change we measure is not a result of long-term behavioral drift.

Because our design requires tracking behavioral change across multiple treatments, random sampling from the ``garden hose'' is inappropriate. Instead, we identified a specific sub-population of politically-engaged Twitter users and created a large ``computational focus group'' to track their collective behavior over time as a panel~\cite{lin2013focus}. While this limits the generalizability of our findings, it also provides important insights to users within the political domain. If a user tweeted using a hashtag like ``\#debate'' or mentioned one of the candidates' Twitter accounts during any of the four presidential debates and their tweet appeared in the Twitter ``garden hose'' streaming API\footnote{https://dev.twitter.com/docs/streaming-apis}, the user was selected into our user pool. Next, we collected the complete tweeting history for these users going back to mid-August using Twitter's REST API\footnote{https://dev.twitter.com/docs/api/1.1}. Because these queries are expensive owing to rate limits, we prioritized users who tweeted during more of the debates. Thus users who tweeted during all four debates are more likely to be represented in the sample than users who tweeted during only one of the debates. The resulting corpus has 290,119,348 tweets from 193,532 unique users including elites such as politicians, journalists, and pundits as well as non-elite partisans and aspiring comedians.

For each of the eight events, we examined tweets made during a 48-- to 96--hour window covering the event itself and its aftermath. Within these windows, we examined tweet volumes and identified the hour containing the peak level of cumulative activity. All subsequent analyses in the paper will look at aggregate behavior in this one hour window of peak activity. Descriptive statistics for the number for the time of the window, unique users, tweets, retweets, mentions, and hashtags observed in each of the 12 events are summarized in Table~\ref{tab:pop}. A ``political relevance ratio'' was also calculated to validate the differences between political events. This ratio is the fraction of tweets during each of the events that containing the names (\textit{e.g.}, ``Obama'' or ``Romney''), candidates' twitter handles (\textit{e.g.}, ``\@barackobama'' or ``\@mittromney''), or any of the the events (\textit{e.g.}, ``DNC'', ``RNC'', ``debate'', ``benghazi'', ``47 percent'', etc.) at the peak time.

\begin{figure*}[!htb]
\def\fname{vol_per_min}\small
\hspace*{-.8cm}
\begin{tabular}{cccc}
(a) DEB 1 & (b) PRE 1 & (c) CONV 1 (RNC) & (d) NEWS 1 (Benghazi)\\
        \includegraphics[trim=.38cm 0cm .5cm 0cm, clip=true,width=.24\textwidth]{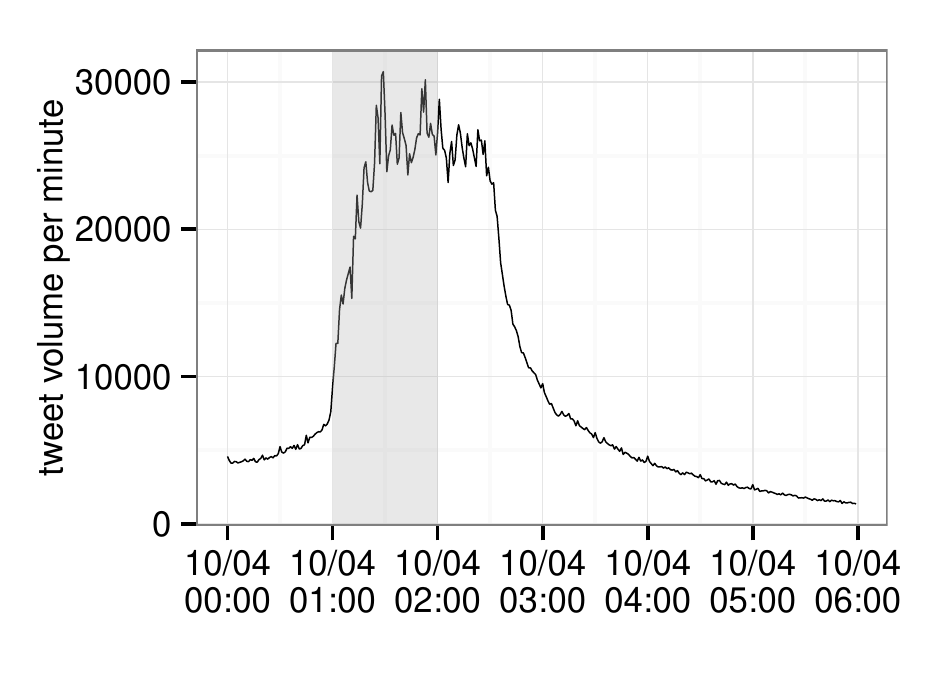} &
        \includegraphics[trim=.38cm 0cm .5cm 0cm, clip=true,width=.24\textwidth]{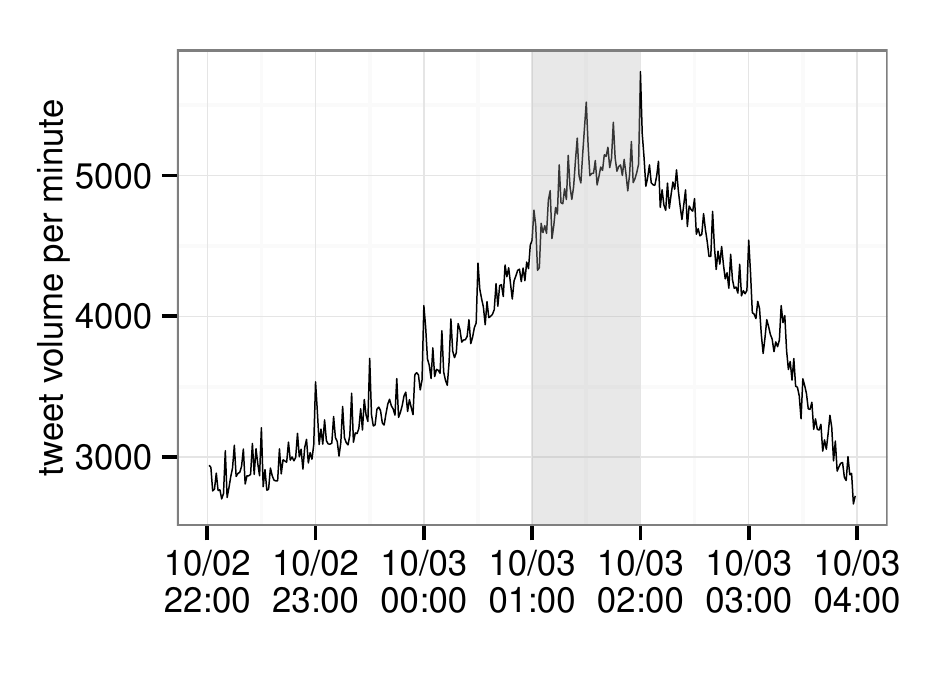} &
        \includegraphics[trim=.38cm 0cm .5cm 0cm, clip=true,width=.24\textwidth]{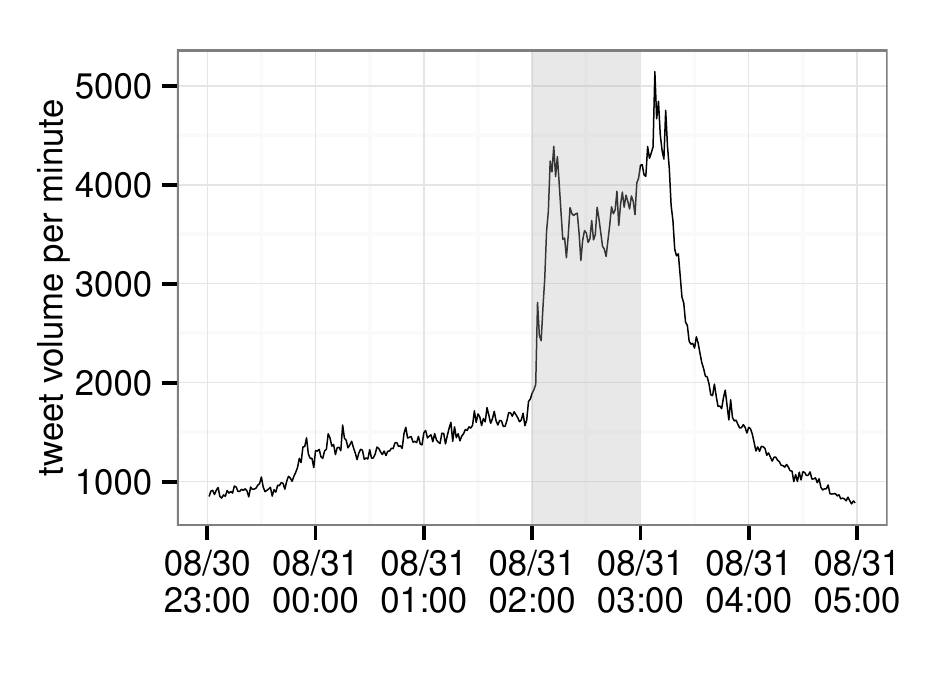} &
        \includegraphics[trim=.38cm 0cm .5cm 0cm, clip=true,width=.24\textwidth]{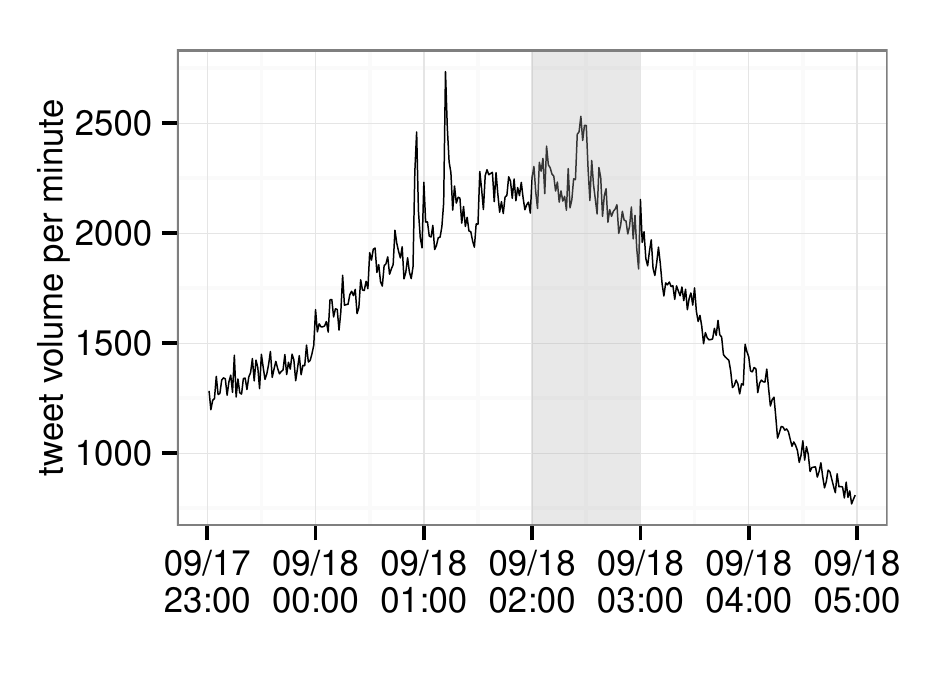}\\
\end{tabular}
\vspace{-2em}
    \caption{%
Tweet volume per minute. Number of tweets per minute in four of the 12 datasets. (a) The six hours during the first debate event (``DEB''). For other categories, we plot the six hour volume centering around the peak within the data range: (b) Normal period prior to the first debate evening (``PRE''). (c) Republican National Convention (RNC)  event (``CONV''). (d) Breaking political news event regarding Benghazi attack (``NEWS'').
        \label{fig:vol_per_min}}
\end{figure*}

Figure~\ref{fig:vol_per_min} plots the distributions of tweet volumes for the hours preceding and following the one-hour window we analyzed. We show one instance for each category. The peaks for the debates correspond to the first hour of each debate, the peak for the RNC corresponds to Clint Eastwood's famous keynote directed at an empty chair at 21:00 EDT on August 30, the peak for the Benghazi attacks corresponds to 20:00 EDT on September 11 when the first American fatality was reported. 

\subsection{Features of communication}
Tweets encode a variety of relational features such as hashtags, mentions, replies, and retweets. Using these features, we construct networks to measure the distributions of users', hashtags', and tweets' centralities.
\begin{itemize}
    \item \textbf{Hashtags} (\#) -- This user-to-hashtag network models when user Alice mentions hashtag ``\#foo'' in a tweet. Alice's hashtag out-degree reflects the number of hashtags she has referenced and the hashtag in-degree of ``\#foo'' reflects the number of users referencing it.
    \item \textbf{Mentions} (@) -- This user-to-user network models when user Alice addresses user Bob anywhere in her tweet. This notifies Bob and lets anyone following her see the tweet. Alice's mention out-degree reflects the number of unique users she has mentioned in her tweets and Bob's mention in-degree reflects the number of unique users who have mentioned him in their tweets.
    \item \textbf{Replies} -- This user-to-user network models when user Alice addresses user Bob as the start of her tweet. This notifies Bob and only lets the users following both of them see the tweet. Alice's reply out-degree reflects the number of unique users she has replied to in her tweets and Bob's reply in-degree reflects the number of unique users who have replied to him in their tweets.
    \item \textbf{Retweets (RT)} -- This user-to-tweet network models when user Alice repeats a tweet from user Bob. Alice's retweet out-degree reflects the number of tweets she has re-tweeted and Bob's tweet in-degree reflects the number of users who have retweeted that tweet. The in-degrees of all of Bob's tweets can be summed to make a \textbf{Retweet user in-degree (RT user)} reflecting the number of times all of Bob's tweets have been retweeted by all other users.
\end{itemize}

\section{Results}\label{sec:results}
We review the results of our design by comparing the changes in communication patterns across all four types of events. We unpack these differences in activity levels across event types by analyzing the changes in the rate of adopting novel content for each type of event. To understand whether these differences are the result distributed activity across all users or the concentration of activity around a few users, we examine whether the distributions of activity for media events differ significantly from news events and the pre-event baselines. Finally, we analyze the relationships between users' audience size and their positions in these activity networks. 

\subsection{Changes in communication}\label{sec:change_comm}
\begin{figure*}[!htb]
\hspace*{-.8cm}
\begin{tabular}{ccccc}
(a) tweet volume & (b) hashtag ratio & (c) mention ratio & (d) reply ratio & (e) RT ratio \\
        \includegraphics[trim=.38cm 0cm .5cm 0cm, clip=true,width=.204\textwidth]{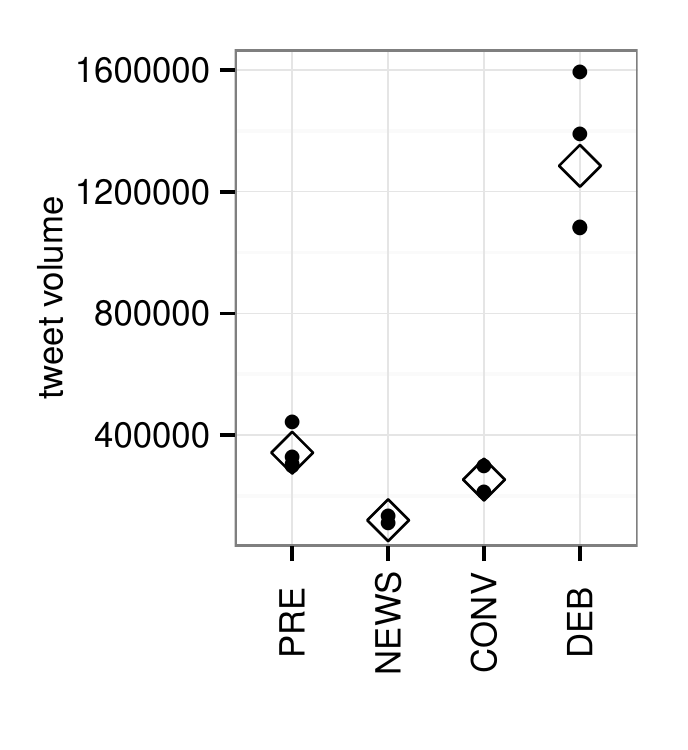} &
        \includegraphics[trim=.38cm 0cm .5cm 0cm, clip=true,width=.186\textwidth]{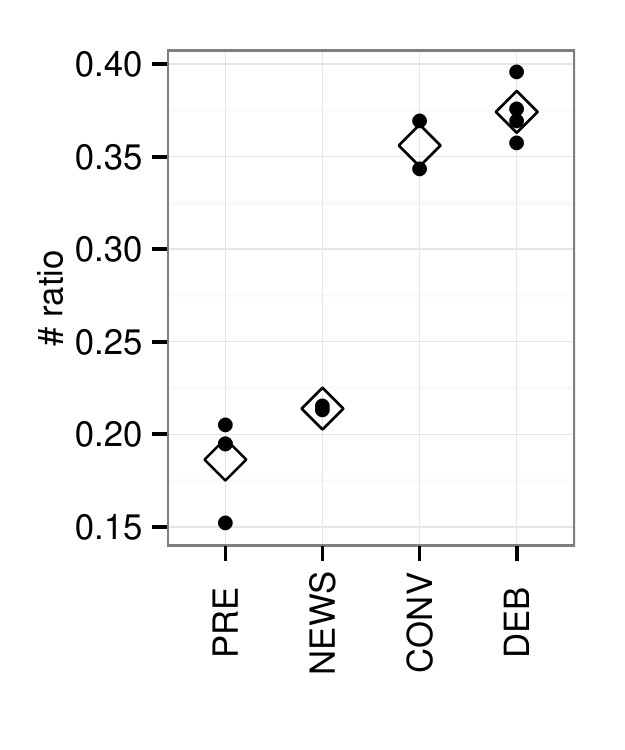} &
        \includegraphics[trim=.38cm 0cm .5cm 0cm, clip=true,width=.186\textwidth]{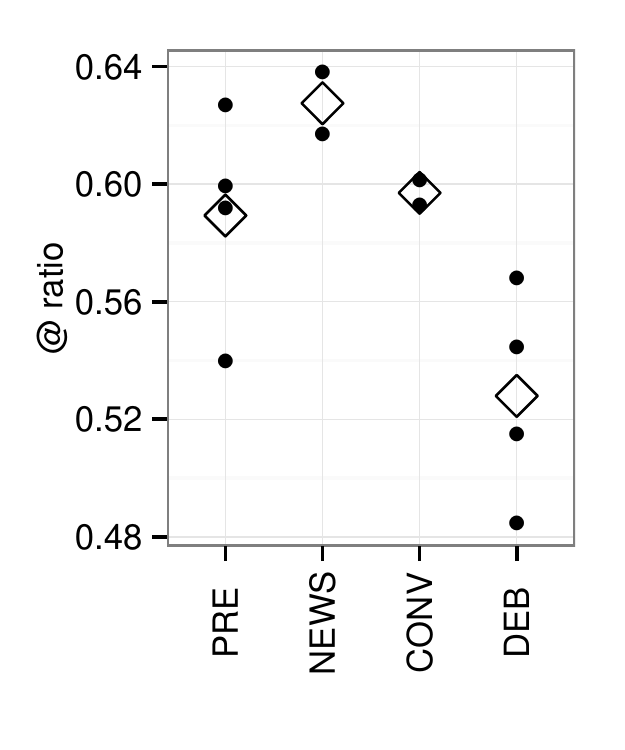} &
        \includegraphics[trim=.38cm 0cm .5cm 0cm, clip=true,width=.186\textwidth]{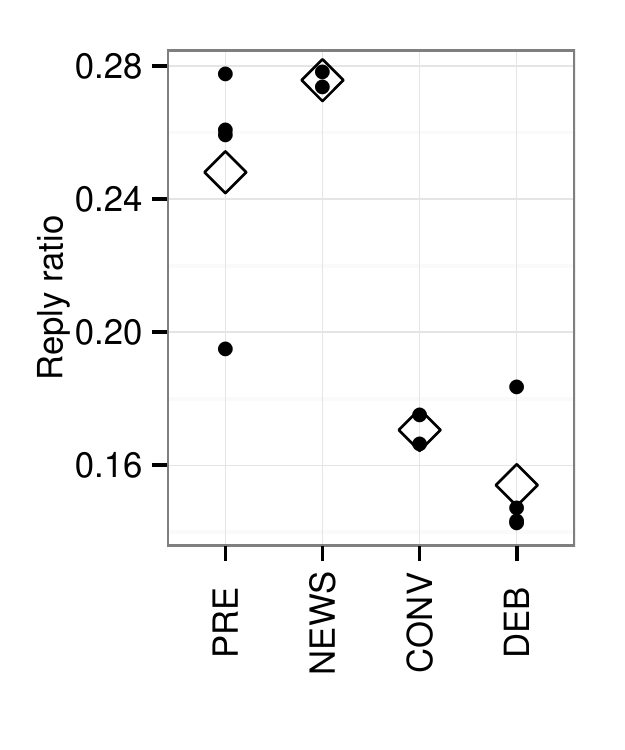} &
        \includegraphics[trim=.38cm 0cm .5cm 0cm, clip=true,width=.186\textwidth]{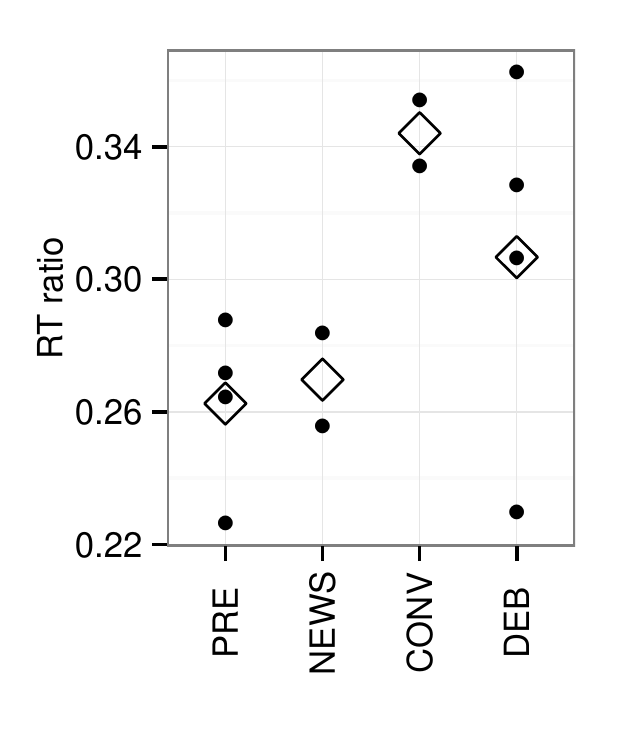} \\
\end{tabular}
\vspace{-2em}
    \caption{%
Changes in communication volume. Twitter communication volume change during the peak time in different events. Diamond shapes indicate the mean value of each category. (a) The tweet volumes at the peak hour in the 12 events. (b) The ratio of tweets with at least one hashtag to the total tweets at the peak hour. (c) The ratio of tweets mentioning a user to the total tweets at the peak hour. (d) The ratio of tweets replying to users to the total tweets at the peak hour. (e) The ratio of retweets to the total tweets at the peak hour.
        \label{fig:ratio}}
\end{figure*}

Figure~\ref{fig:ratio} plots the changes in communication volumes for the twelve observations grouped by event type (typical, news event, national convention, and debate). Results show an increase in topical communication and a decrease in inter-personal communication during shared attention. Tweet volumes for the debates are three to four times greater than during the other types of events (\ref{fig:ratio}a) and the rate of hashtag use nearly doubles during media events over the non-media event rate (\ref{fig:ratio}b). Because hashtags are an \emph{ad hoc} way to create a sub-community by affiliating a tweet with a label others are using, the rise of this behavior during media events suggests users are broadcasting diffuse interests in topics rather than engaging in interpersonal discussions. 

The ratio of tweets that include any mentions of users in the tweet (\ref{fig:ratio}c) and the fraction of tweets that were replies to one or more users (\ref{fig:ratio}d) declines substantially during media events like the debates. This 40\% decline in directed communication suggests media events may not only dominate attention but also change social media behavior to become less interpersonal and more declarative. Finally, the ratio of retweets among all tweets (\ref{fig:ratio}e) increases 20\% over the typical and news events during the more media event-like convention and debates even though these occurred at different times and with different levels volumes of activity.

\begin{figure*}[!htb]
\def\fname{PK_user2minutes}
\hspace*{-.8cm}
\begin{tabular}{ccccc}
(a) Adopt hashtags & (b) Adopt mentions & (c) Reply users & (d) Adopt RTs & (e) Adopt users' RT \\
        \includegraphics[trim=.4cm 0cm .5cm 0cm, clip=true,width=.19\textwidth]{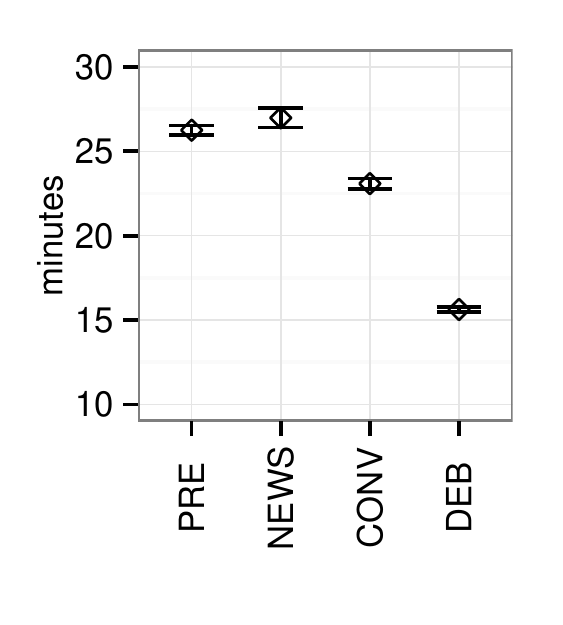} & 
        \includegraphics[trim=.4cm 0cm .5cm 0cm, clip=true,width=.19\textwidth]{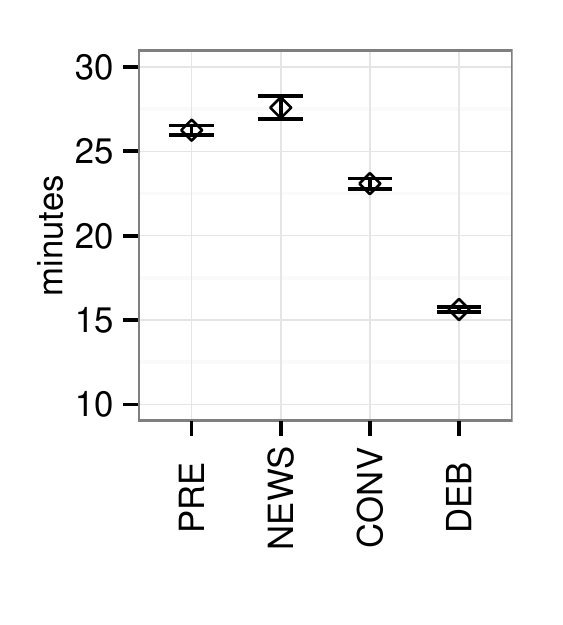} &
        \includegraphics[trim=.4cm 0cm .5cm 0cm, clip=true,width=.19\textwidth]{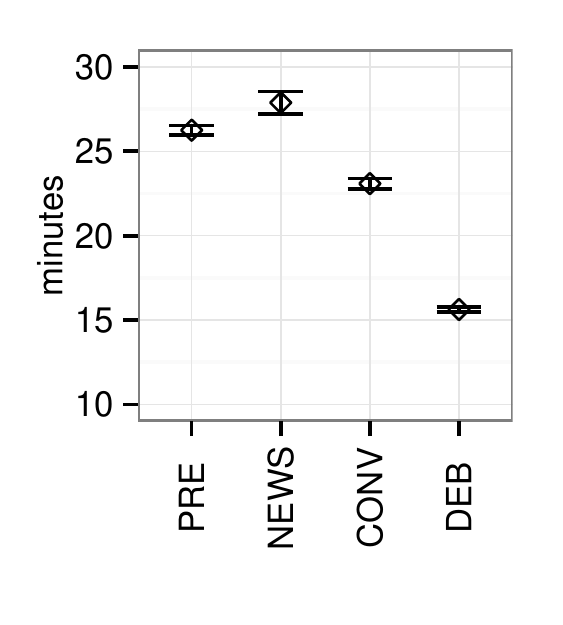} &
        \includegraphics[trim=.4cm 0cm .5cm 0cm, clip=true,width=.19\textwidth]{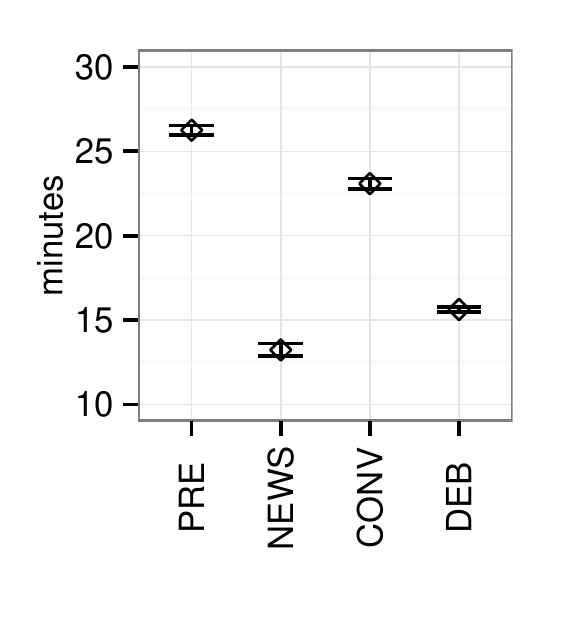} &
        \includegraphics[trim=.4cm 0cm .5cm 0cm, clip=true,width=.19\textwidth]{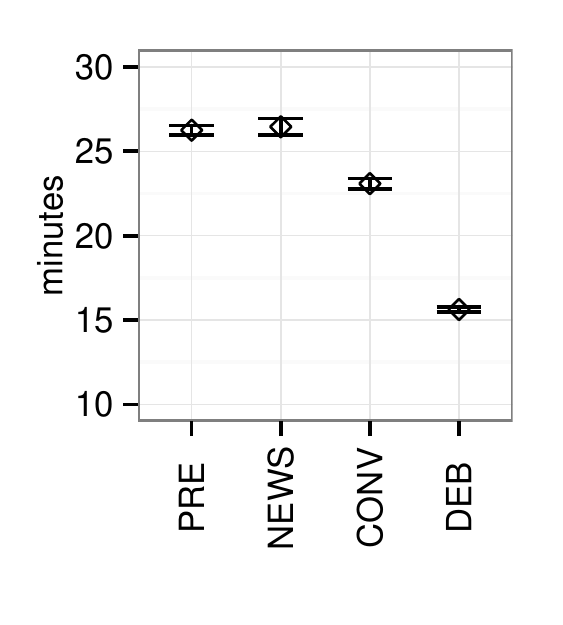} \\
\end{tabular}
\vspace{-2em}
    \caption{%
Adoption lags for different event types. Average lag time per user (in minutes) to adopt novel items (hashtags, mentions, users, retweets).
        \label{fig:lag}}
\end{figure*}

\subsection{Changes in concentration}\label{sec:change_sys}

\begin{figure*}[!htb]
\def\fname{PK_deg_distr}
\def\gname{PK_deg_distr_gini}
\def\fext{}
\hspace*{-.7cm}
\begin{tabular}{cccc}
(a) Hashtag in-degree & (b) Mention in-degree & (c) Reply in-degree & (d) RT in-degree\\
        \includegraphics[trim=.7cm 0cm .36cm 0cm, width=.23\textwidth]{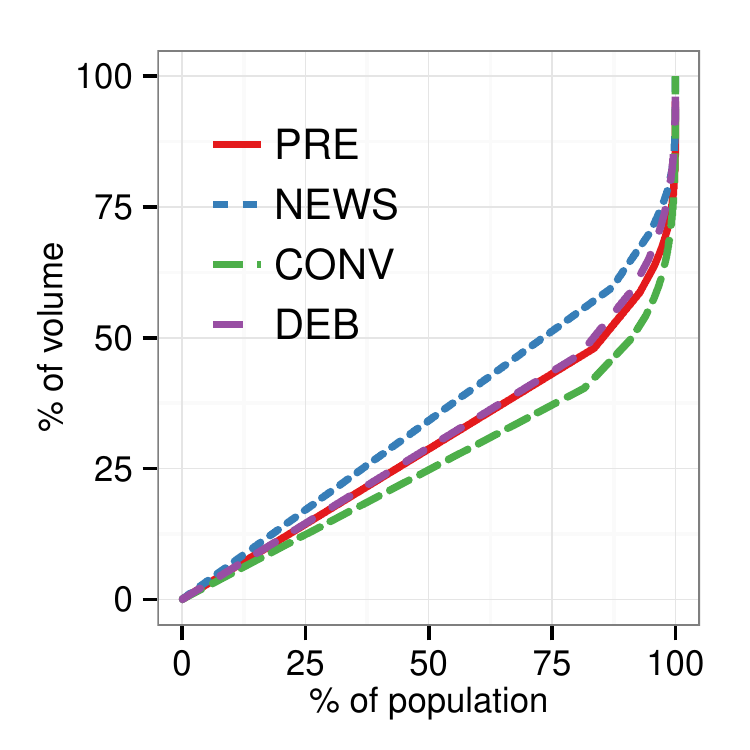} &
        \includegraphics[trim=.7cm 0cm .36cm 0cm, width=.23\textwidth]{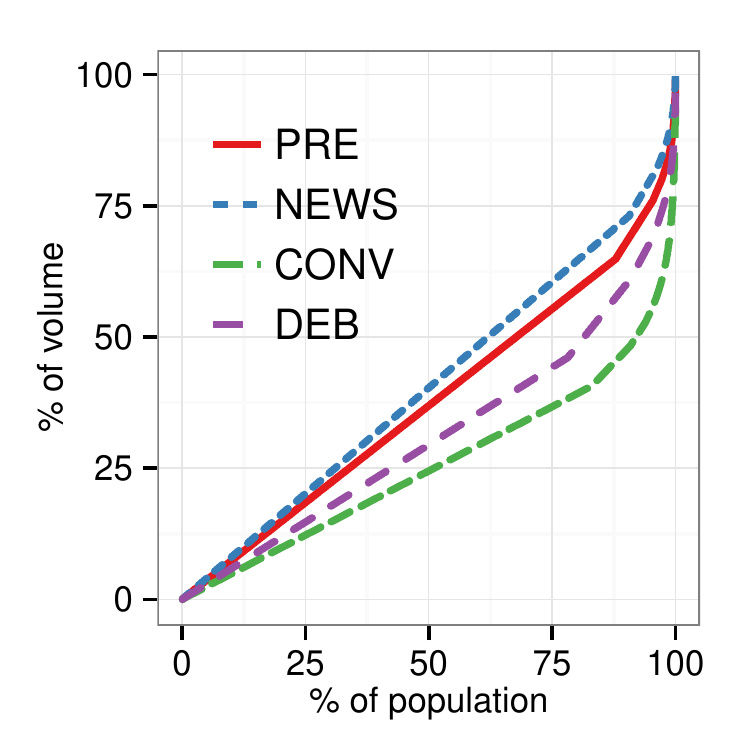} & 
        \includegraphics[trim=.7cm 0cm .36cm 0cm, width=.23\textwidth]{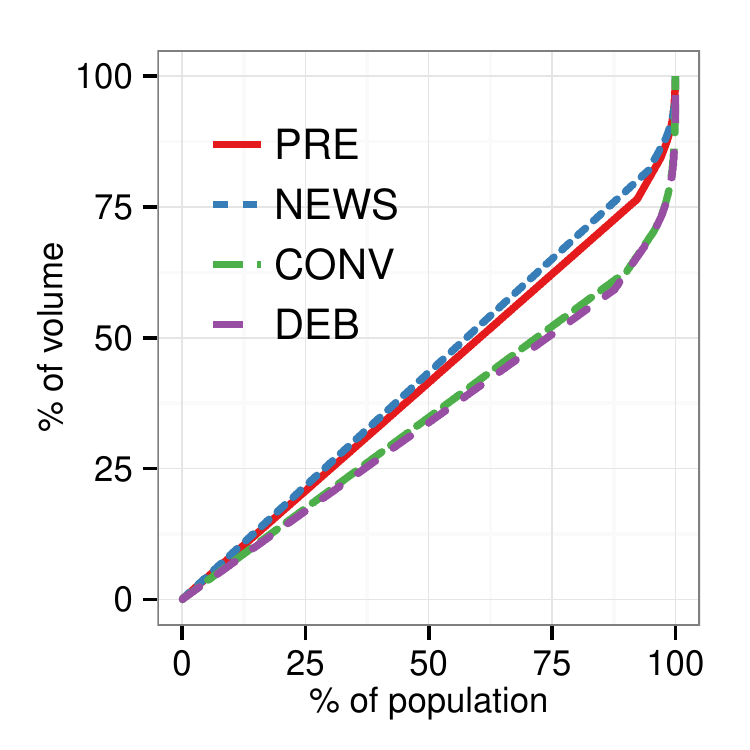} &
        \includegraphics[trim=.7cm 0cm .36cm 0cm, width=.23\textwidth]{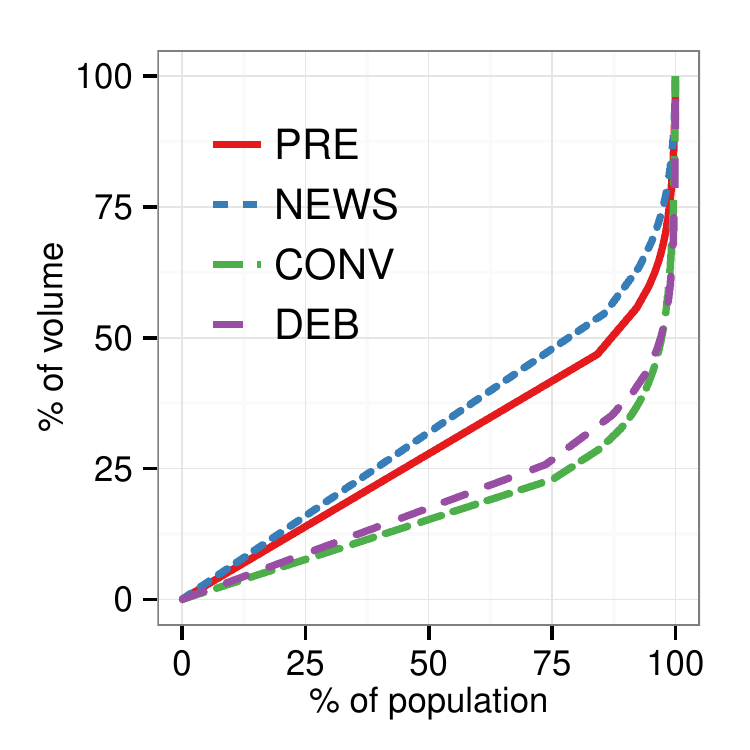} \\
        
(e) Hashtag out-degree & (f) Mention out-degree & (g) Reply out-degree & (h) RT out-degree\\
        \includegraphics[trim=.7cm 0cm .36cm 0cm, width=.23\textwidth]{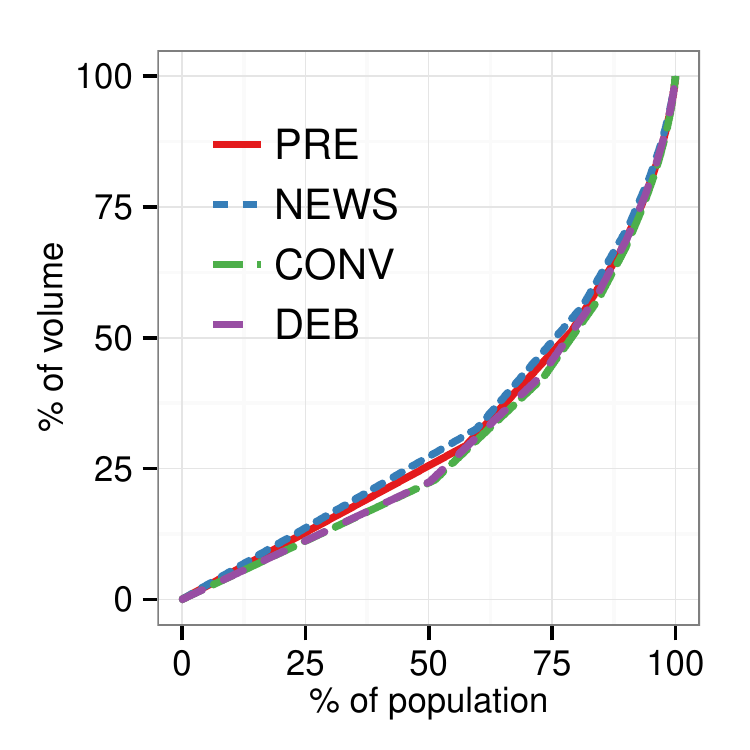} &
        \includegraphics[trim=.7cm 0cm .36cm 0cm, width=.23\textwidth]{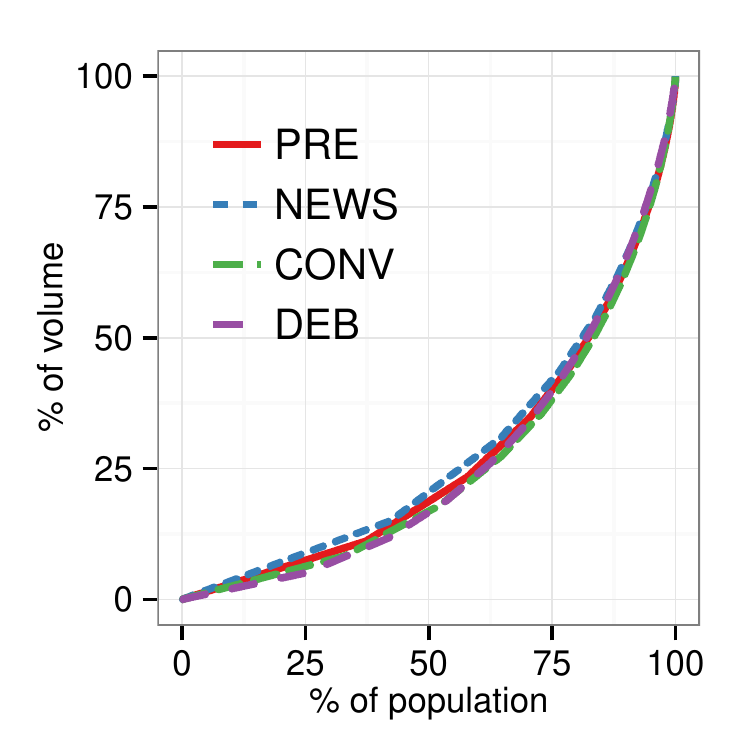} &
        \includegraphics[trim=.7cm 0cm .36cm 0cm, width=.23\textwidth]{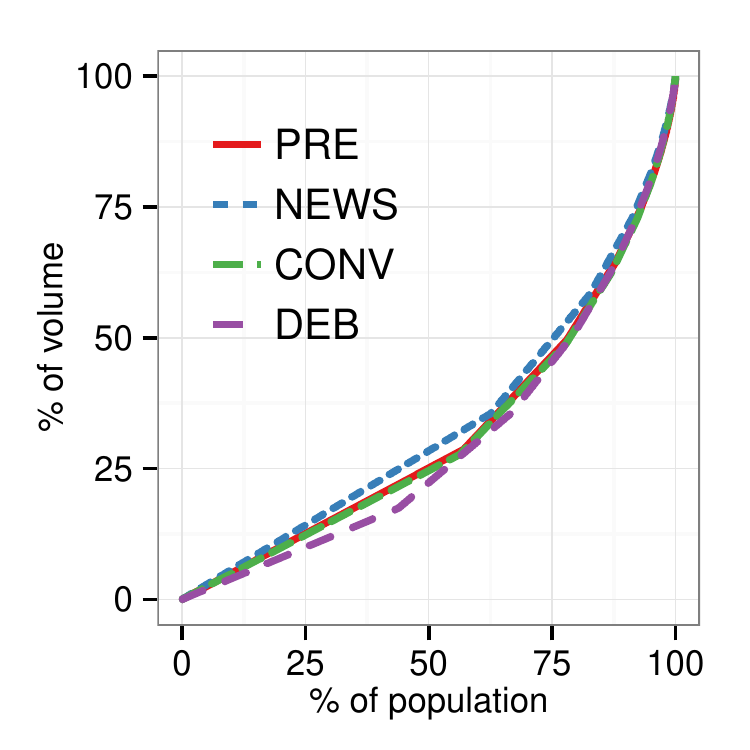}&
        \includegraphics[trim=.7cm 0cm .36cm 0cm, width=.23\textwidth]{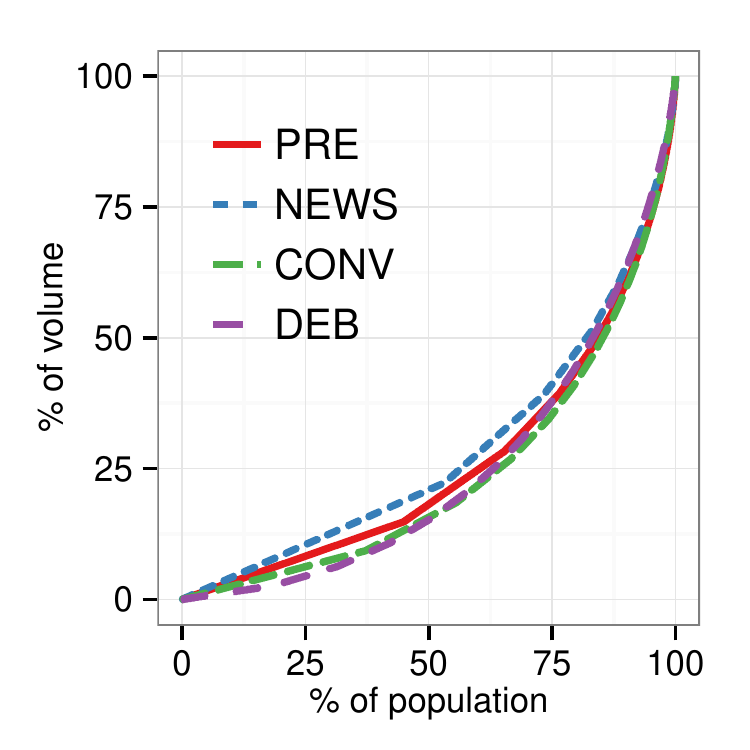} \\

\end{tabular}
\vspace{-1em}
    \caption{%
Lorentz curves for cumulative degree distributions of activity. Increasing equality converges toward diagonal line from the origin to the upper-right and increasing inequality converges toward a hyperbola rising to 100\% of volume at the 100th percentile. 
        \label{fig:deg}}
\end{figure*}
The previous sections demonstrated significant changes in the aggregate behavior of the users' population . However it is unclear whether these differences are driven by most users becoming more receptive to elite messaging (``rising stars'') or if all users are becoming more engaged (``rising tides''). 

We use Lorenz curves to plot the cumulative distribution of activity in the system for each of the four types of events. A Lorenz curve shows for the bottom $x\%$ of users or tweets, the percentage $y\%$ of the activity they generated. Thus more equally-distributed activity will be more linear while more highly concentrated activity will more parabolic. Relative to the typical pre-debate events, a top-down pattern of ``rising stars'' should be indicated by a concentration of activity in a few users while a bottom-up pattern ``rising tides'' should be indicated by a flattened distributions.

We measure the level of degree concentration in these Lorenz curves using the Gini coefficient. It is defined as the ratio of the area that lies between the line of equality (the line at 45 degrees) and the Lorenz curve over the total area under the line of equality. The Gini coefficient for a set of users or tweets P with degrees $y_i$ ($i=1,...,n$) and probability function $f(y_i)$ is given by:
\[
G = 1-\frac{\sum_{i=1}^n f(y_i) (S_{i-1}+S_i)}{S_n},
\]
where $S_i=\sum_{j=1}^i f(y_j)y_j$ and $S_0=0$. The Gini coefficient is a measure for identifying preferential patterns in general, as opposed to measures such as power-law exponent which can only apply to networks following power-law distribution.

Figure~\ref{fig:deg} plots the in- and out-degree Lorenz curves for the four networks of hashtags, mentions, replies, and retweets. The out-degrees for all four types of activity networks show significant similarities across the four event types and comparatively high levels of concentrated activity: the top users are responsible for most hashtag references (\ref{fig:deg}e), mentioning other users (\ref{fig:deg}f), replying to other users (\ref{fig:deg}g), and retweeting users' content (\ref{fig:deg}h). These findings together suggest the concentration of users' attention to content in the networks is very similar regardless of media events. 

The in-degrees show more substantial differences between event types as well as truncated distributions. The top 20\% of hashtags make up more than 50\% of all observed hashtags (\ref{fig:deg}a), with this concentration being more exaggerated around the national conventions and more relaxed around the news events. This reflects the dominance of a official hashtags like ``\#nbcpolitics'' during the conventions and the presence of many unrelated but popular hashtags like ``\#moviesyoucantdislike'' during the unscheduled news events. The convention and debate media events also drove increased concentration of mention (\ref{fig:deg}b) and reply activity (\ref{fig:deg}c) around top users as compared to pre-events and news events. This is suggestive of many users directing their tweets towards elite users rather than each other. Finally, retweet activity (\ref{fig:deg}d) on the six media events shows the largest differences from the pre-debate baseline (see Table~\ref{tab:ks}). The top 25\% of users' tweets accounted for approximately 75\% of all retweet activity, clear evidence that users' behavior under conditions of shared attention become increasingly concentrated around elites rather than increasingly distributed across many users.

In addition to the differences \textit{within} the distributions of for a given type of event, the Lorenz curves in Figure \ref{fig:deg} are also qualitatively different across the in-degree versus out-degree distributions: out-degrees show a pattern of consistent inequality throughout the population while in-degrees show a pattern of even distribution until rising sharply at the 90$^{\textit{th}}$ percentile. Figure \ref{fig:deg_vs_gini} also captures these differences in in-degree versus out-degree concentration: in-degree concentrations ($y$-axis) vary considerably across event types while out-degree concentrations are consistent across event types. We discuss this observation in more detail in the following section.

To test whether the concentration of activity differed significantly from the pre-debate event baselines, we measured the deviation of each event type's Lorenz curve from the pre-debate baseline events using a two-sample Kolmogorov-Smirnov (K-S) test~\cite{gail1978scale}. The K-S statistics in Table~\ref{tab:ks} confirm the Lorenz curves for networks of mentions, replies and retweets are significantly different during the news events, national conventions, and debates. The statistics for the differences between the out-degrees are generally much larger than the statistics for the in-degrees, reflecting the larger differences between these curves discussed above. Taken together, users reproduce similar \textit{behaviors} of focusing their activity on certain tweets, users, and hashtags across all types of events even though the tweets, users, and hashtags that are the \textit{focus} of this attention becomes more concentrated during media events.

\begin{table}[!bht]
    \centering \small
    \begin{tabular}{ rl lllll }
    \toprule              
& & Hashtag & Mention & Reply & RT user \\
    \midrule              
& & & & & \\
\multirow{3}{*}{\vrtLbl{in-deg.}{-3}} & NEWS & 0.133 & 0.074 & 0.040 & 0.108 \\
&CONV & 0.113 & 0.280 & 0.145 & 0.339 \\
&DEB & 0.019$^{a}$ & 0.158 & 0.162& 0.274 \\
    \midrule              
& & & & & \\
\multirow{3}{*}{\vrtLbl{out-deg.}{-3}} & NEWS & 0.062 & 0.060&0.067&0.095\\
& CONV & 0.066 &0.052 &0.011$^{b}$&0.091\\
& DEB & 0.037 & 0.083 &0.091&0.125\\
    \bottomrule
    \end{tabular}
    \caption{Kolmogorov-Smirnov test (K-S test) for comparing the PRE curves with the remaining three curves in Figure~\ref{fig:deg}. All the statistics $D$ listed here have p-values $p < 10^{-6}$ unless reported otherwise: $^{a}p = 0.0464$, $^{b}p = 0.592$ (n.s.).}
    \label{tab:ks}
\end{table}

\subsection{Patterns of connectivity and concentration}

\begin{figure*}[!htb]
\def\fname{PK_deg_vs_gini}
\hspace*{-.8cm}
\begin{tabular}{cccccc}
\multirow{5}{*}{\vrtLbl{in-degree}{6}} & (a) Hashtag & (b) Mention & (c) Reply & (d) RT & (e) RT user \\
&
        \includegraphics[trim=.4cm 0cm .5cm 0cm, width=.19\textwidth]{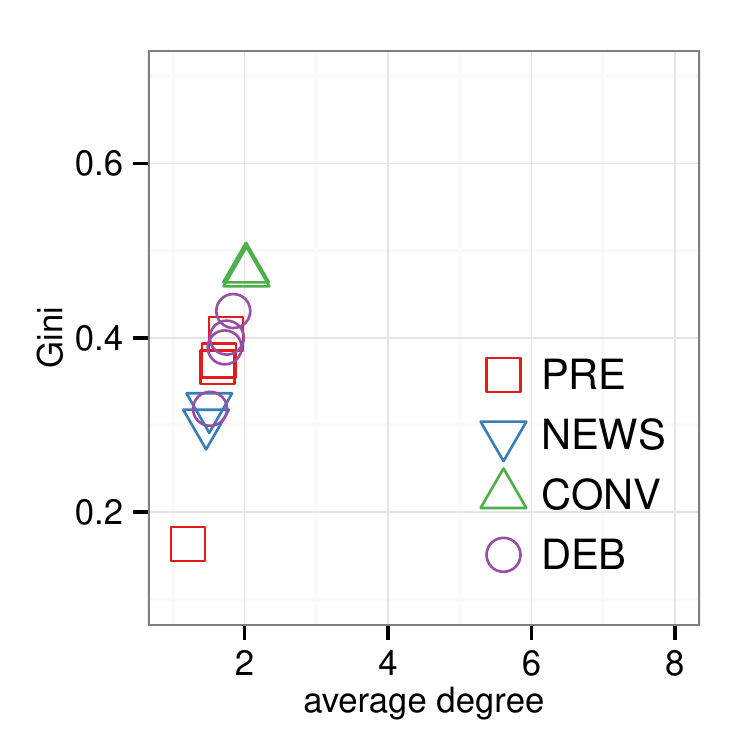} &
        \includegraphics[trim=.4cm 0cm .5cm 0cm, width=.19\textwidth]{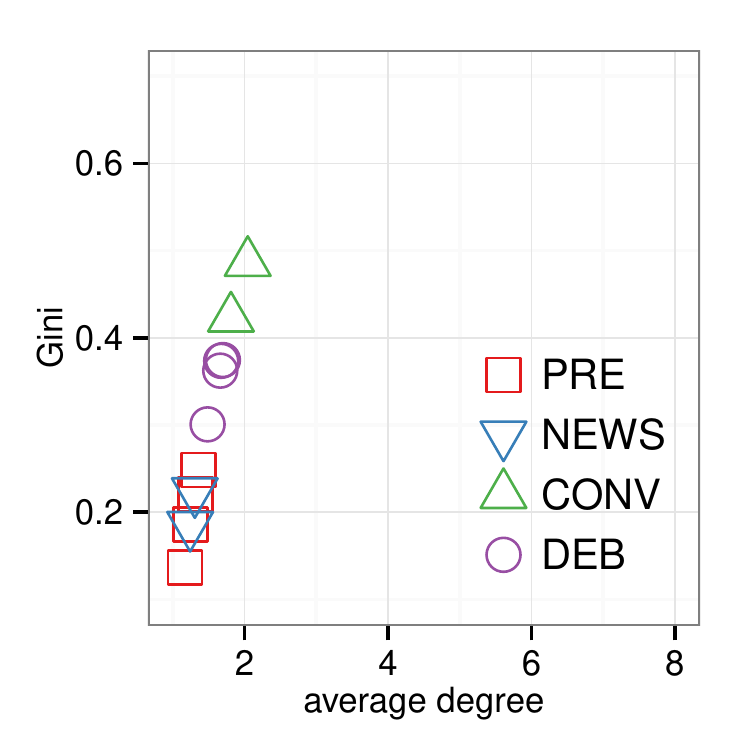}&
        \includegraphics[trim=.4cm 0cm .5cm 0cm, width=.19\textwidth]{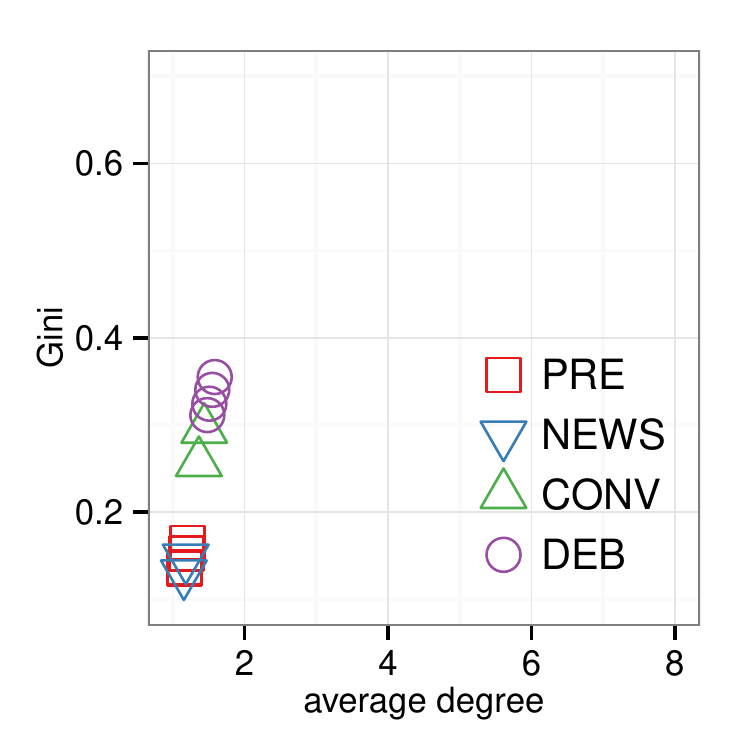} &
        \includegraphics[trim=.4cm 0cm .5cm 0cm, width=.19\textwidth]{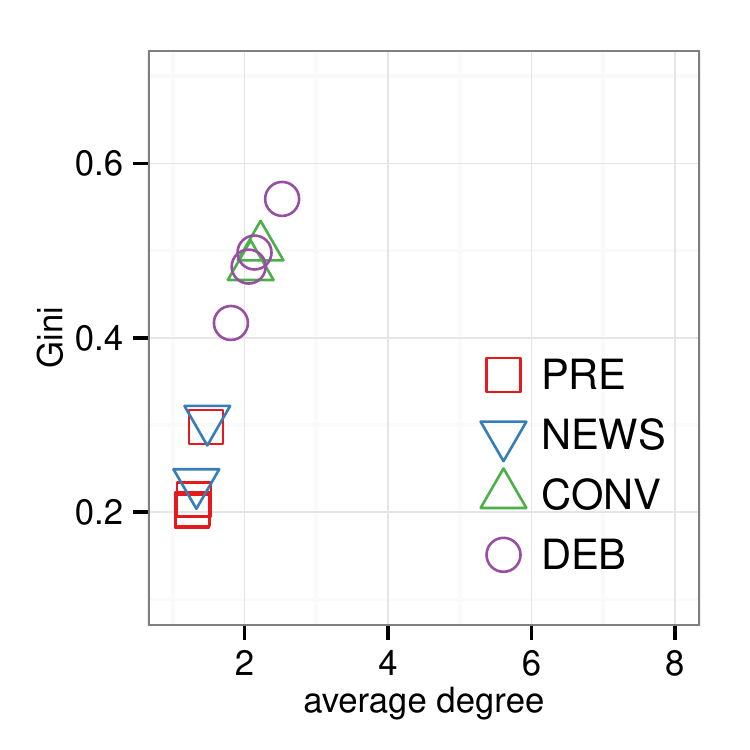} &
        \includegraphics[trim=.4cm 0cm .5cm 0cm, width=.19\textwidth]{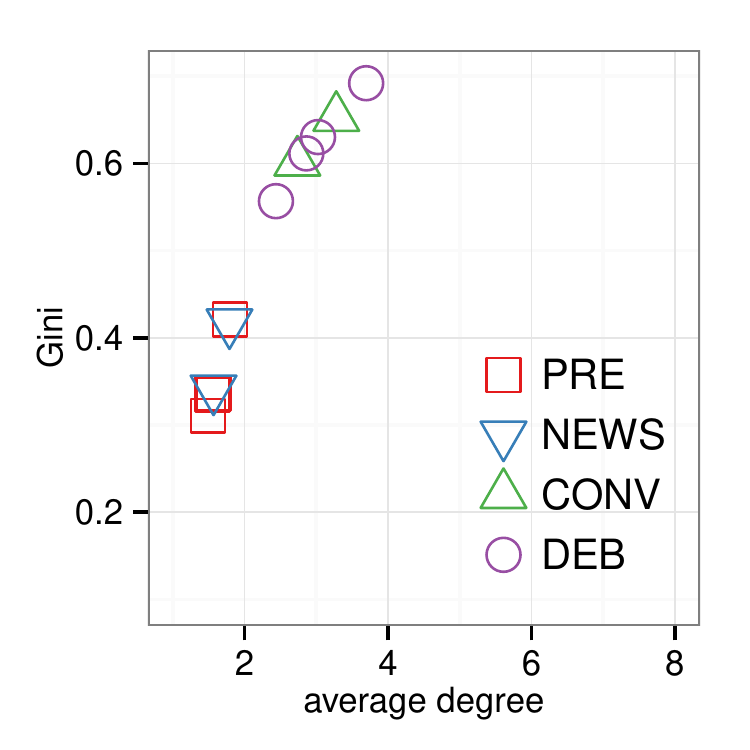} \\
\multirow{5}{*}{\vrtLbl{out-degree}{6}} & (f) Hashtag & (g) Mention & (h) Reply & (i) RT & (j) RT user \\
&
        \includegraphics[trim=.4cm 0cm .5cm 0cm, width=.19\textwidth]{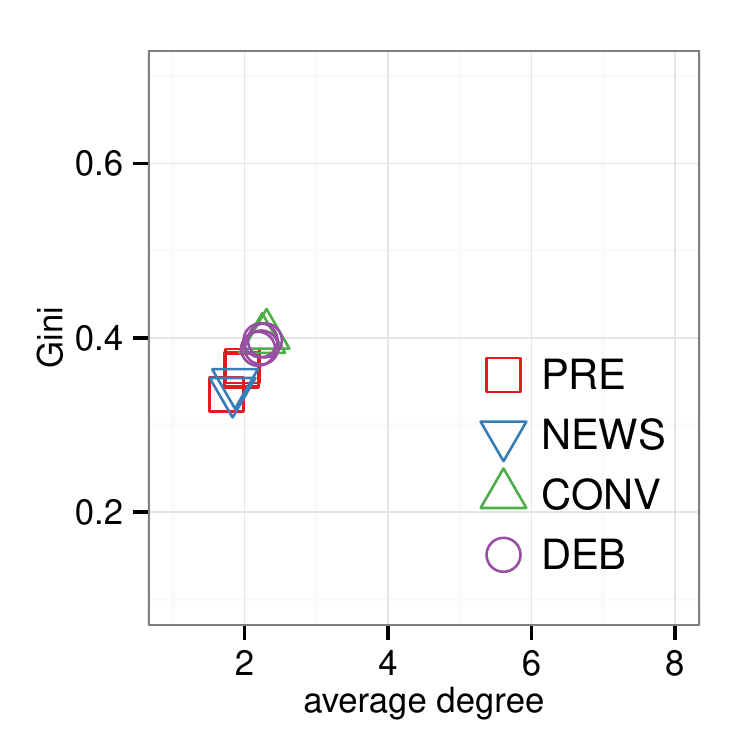} &
        \includegraphics[trim=.4cm 0cm .5cm 0cm, width=.19\textwidth]{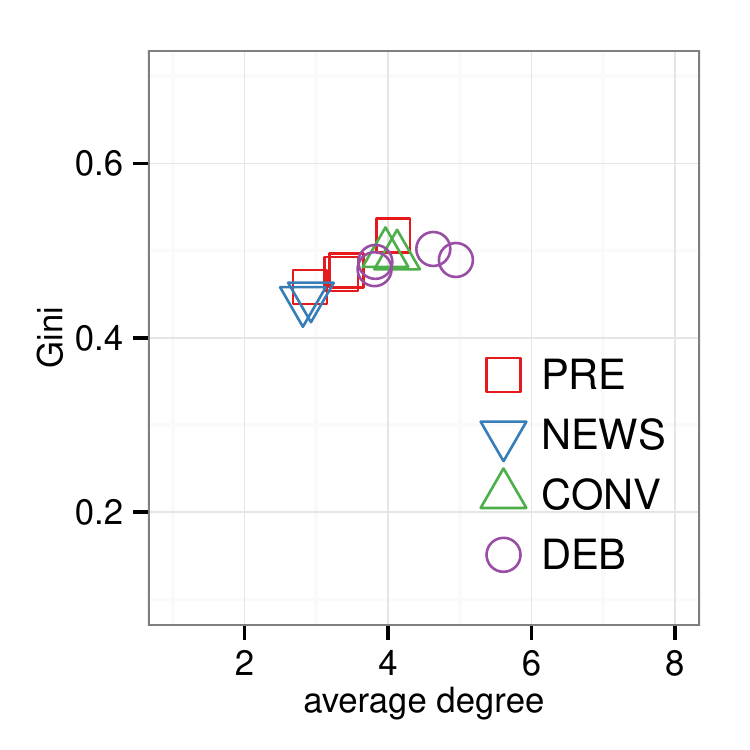}&
        \includegraphics[trim=.4cm 0cm .5cm 0cm, width=.19\textwidth]{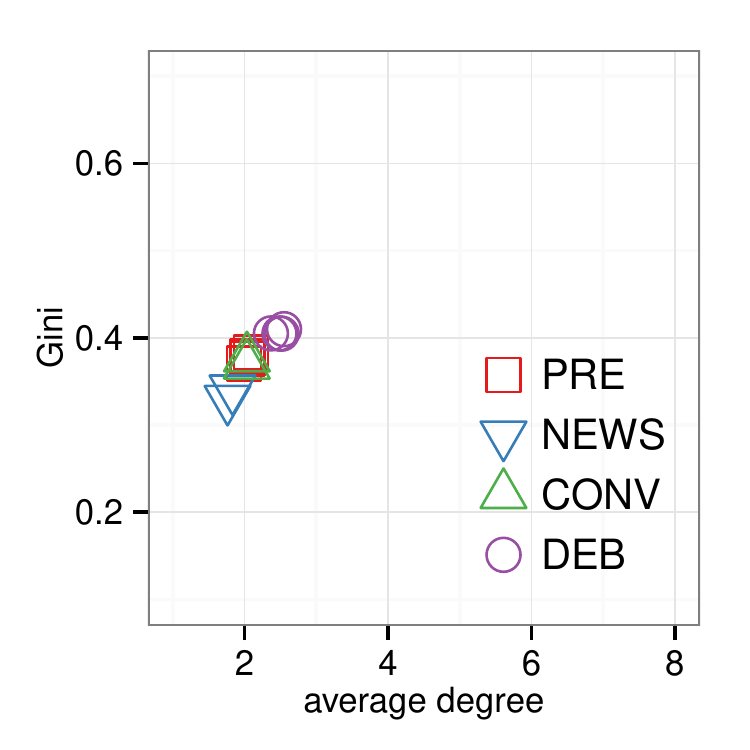} &
        \includegraphics[trim=.4cm 0cm .5cm 0cm, width=.19\textwidth]{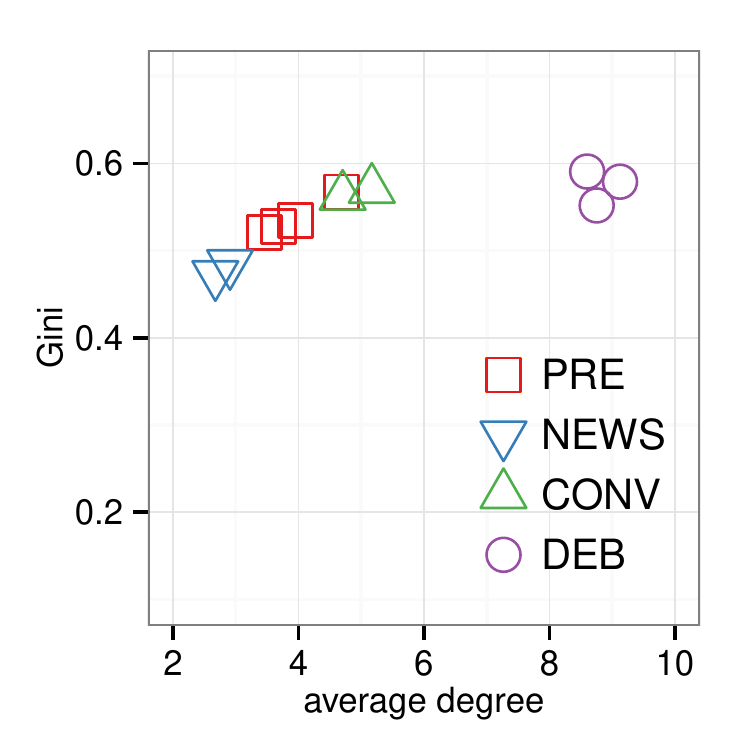} &
        \includegraphics[trim=.4cm 0cm .5cm 0cm, width=.19\textwidth]{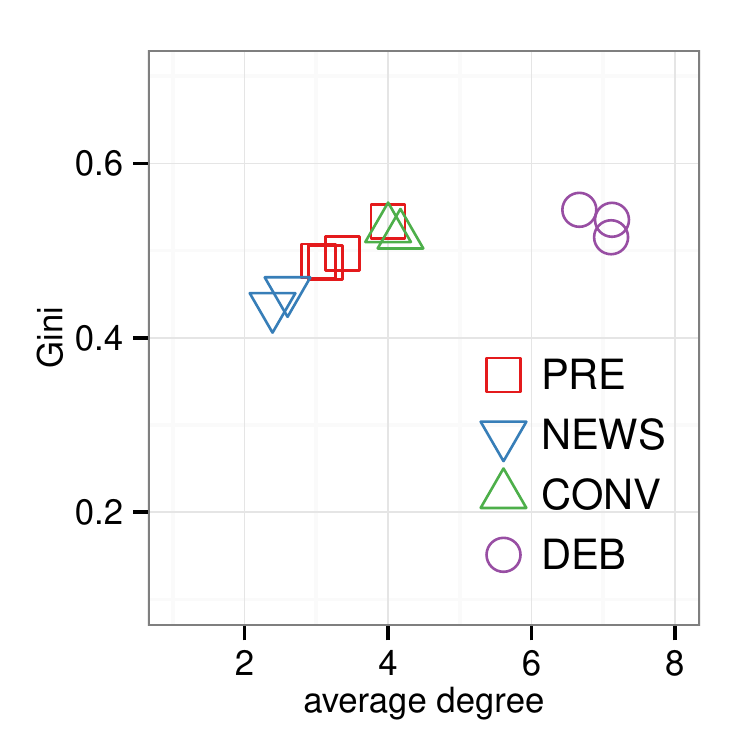} \\
\end{tabular}
\vspace{-1em}
    \caption{%
 Connectivity-concentration state spaces. For each of the twelve observed events, the Gini coefficient for the network's degree distribution is plotted on the $y$-axis and the average degree of the network is plotted on the $x$-axis.
        \label{fig:deg_vs_gini}}
\end{figure*}

To unpack the question of whether the behavioral patterns observed above are driven more by changes across the system of users (``rising tides'') or changes arising from the concentration of activity (``rising stars''), we examine the relationship between these networks' connectivity and concentration. Figure~\ref{fig:deg_vs_gini} shows the Gini coefficient and the average degree of users in the system for each of the activity networks (hashtags, mentions, replies, retweets). System-level changes during media events should be reflected in the increased average degree as more hashtags, users, and tweets become more well-connected ($x$-axis). Alternatively, individual-level changes during media events should be reflected in the increased Gini coefficient as users concentrate their activity around fewer hashtags, users, and tweets ($y$-axis). Networks in which the users are evenly but poorly connected would cluster around the lower-left, networks with poor connectivity but high levels of centralization would cluster in the upper-left, networks with an even distribution of highly connected nodes would cluster in the lower-right, and networks with highly connected but nevertheless highly concentrated activity would cluster in the upper-right. 

The absolute position of each network observation in this space as well as their position relative to networks for other types of events both warrant discussion. In the user-to-hashtag network (\ref{fig:deg_vs_gini}a,\ref{fig:deg_vs_gini}f), all twelve events show moderate amounts of centralization and low levels of connectivity. Users do not use many unique hashtags on average (\ref{fig:deg_vs_gini}a) nor are there more hashtags in circulation on average (\ref{fig:deg_vs_gini}f) during the media events compared to other types of events. There is more centralization among the hashtags than users, reflecting the dominance of official hashtags about events, but the differences between types of events are not major. 

In the user-to-user mention network, media events have a greater tendency for a few users to receive many mentions from other users than other types of events (\ref{fig:deg_vs_gini}b) but most users still receive fewer than two mentions during the peak hour for all types of events. A similar pattern is found in the user-to-user reply network (\ref{fig:deg_vs_gini}c), both of which can be interpreted as changes at the level of the system concentrating activity without corresponding changes at the level of the users increasing connectivity. Returning to the mention network, the mention out-degree spaces show that users mention between three and five other users on average over the course of the peak hours for all events (\ref{fig:deg_vs_gini}g) but much of this mentioning behavior in concentrated in a few, very ``chatty'' users communicating with very many other users. The reply network's out-degree (\ref{fig:deg_vs_gini}h) has half the average connectivity of the mention network's out-degree, indicating users reply to only two other users during peaks across and this effect holds across all types of events.

Retweet behavior shows major differences between event types that were not observed in the other types of activity networks. The in-degrees of retweets for media events in the user-to-retweet network (\ref{fig:deg_vs_gini}d) and user-to-user retweet network (\ref{fig:deg_vs_gini}e) both show limited increases in average connectivity but extremely large increases in centralization. In other words, few users and their tweets ``go viral'' (the average retweeted tweet is only retweeted between one and three times) but there are also users and tweets that are retweeted so widely, they account for a disproportionate amount of observed activity. This pattern on media events differs significantly from the pre-debate baseline and news events which have low connectivity as well as low centralization and demonstrates a stronger shift in the behavior of the system in addition to moderate changes in the behavior of users. 

The out-degree behavior in the user-to-tweet (\ref{fig:deg_vs_gini}i) and user-to-user (\ref{fig:deg_vs_gini}j) retweet networks also show major changes for the debates in particular compared to the baseline and news events. The average user more than doubled their retweeting activity from approximately 4 retweets and 3 users during a peak hour to between 8 and 9 tweets (\ref{fig:deg_vs_gini}i) and 6 and 7 users (\ref{fig:deg_vs_gini}j)  during the debates. This retweet activity is consistently and highly centralized across all event types suggesting some users' focus predominantly on actively retweeting other content.

Across these activity types, the in-degrees saw consistent patterns of increasing centralization rather than degree in response to media events while the out-degrees saw patterns of increasing degree rather than concentration in response to media events. Taken together, this suggests that while users across the system become more active during media events, this additional activity predominately benefits a handful of users, hashtags, and tweets.

\subsection{Changing user responsiveness}

\begin{figure*}[!htb]
\centering
\def\fnameA{scatter_follower}
\hspace*{-.8cm}
\begin{tabular}{ccc}
(a) & (b) & (c)\\
        \includegraphics[trim=.8cm 0cm .5cm 0cm, width=.24\textwidth]{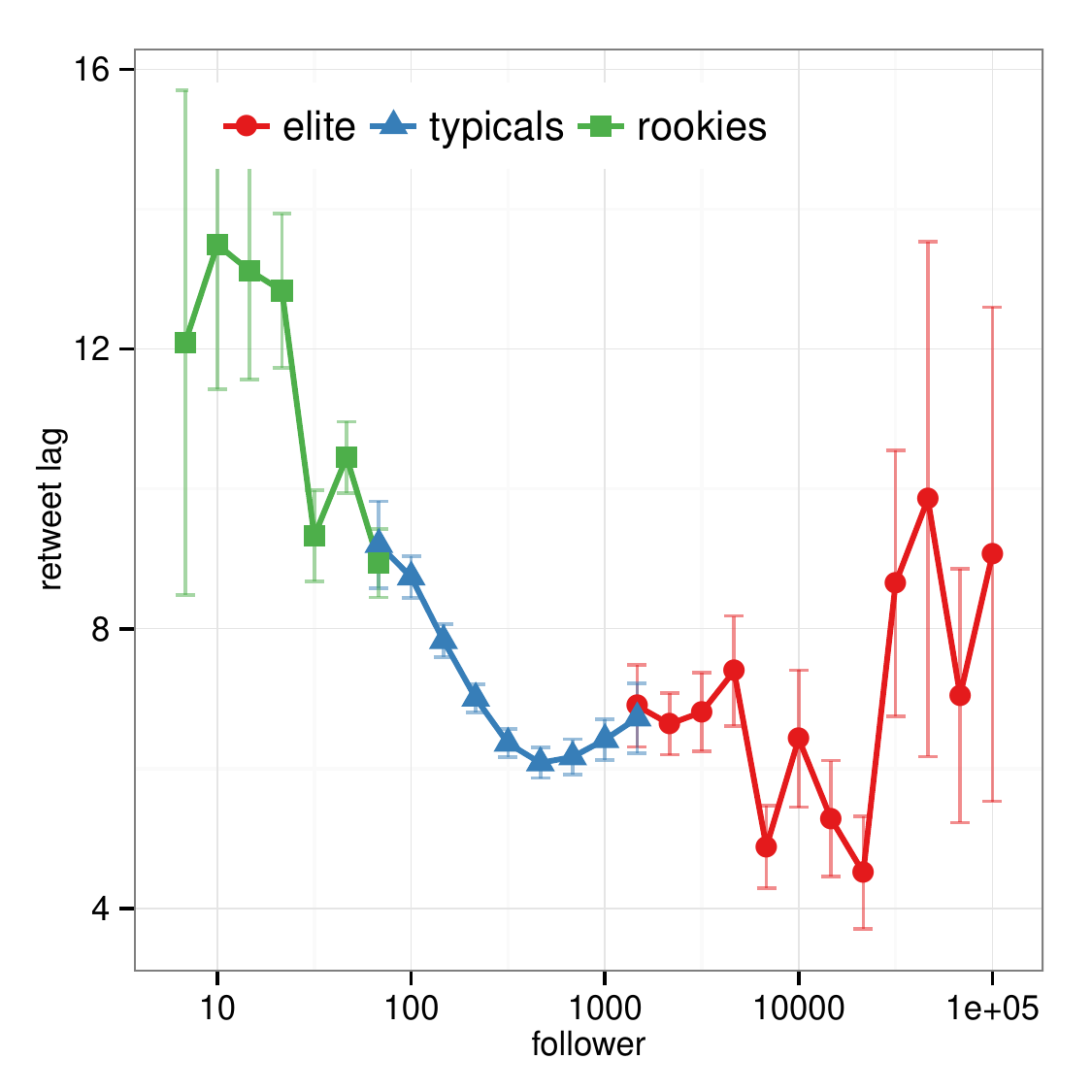} &
        \includegraphics[trim=.8cm 0cm .5cm 0cm, width=.24\textwidth]{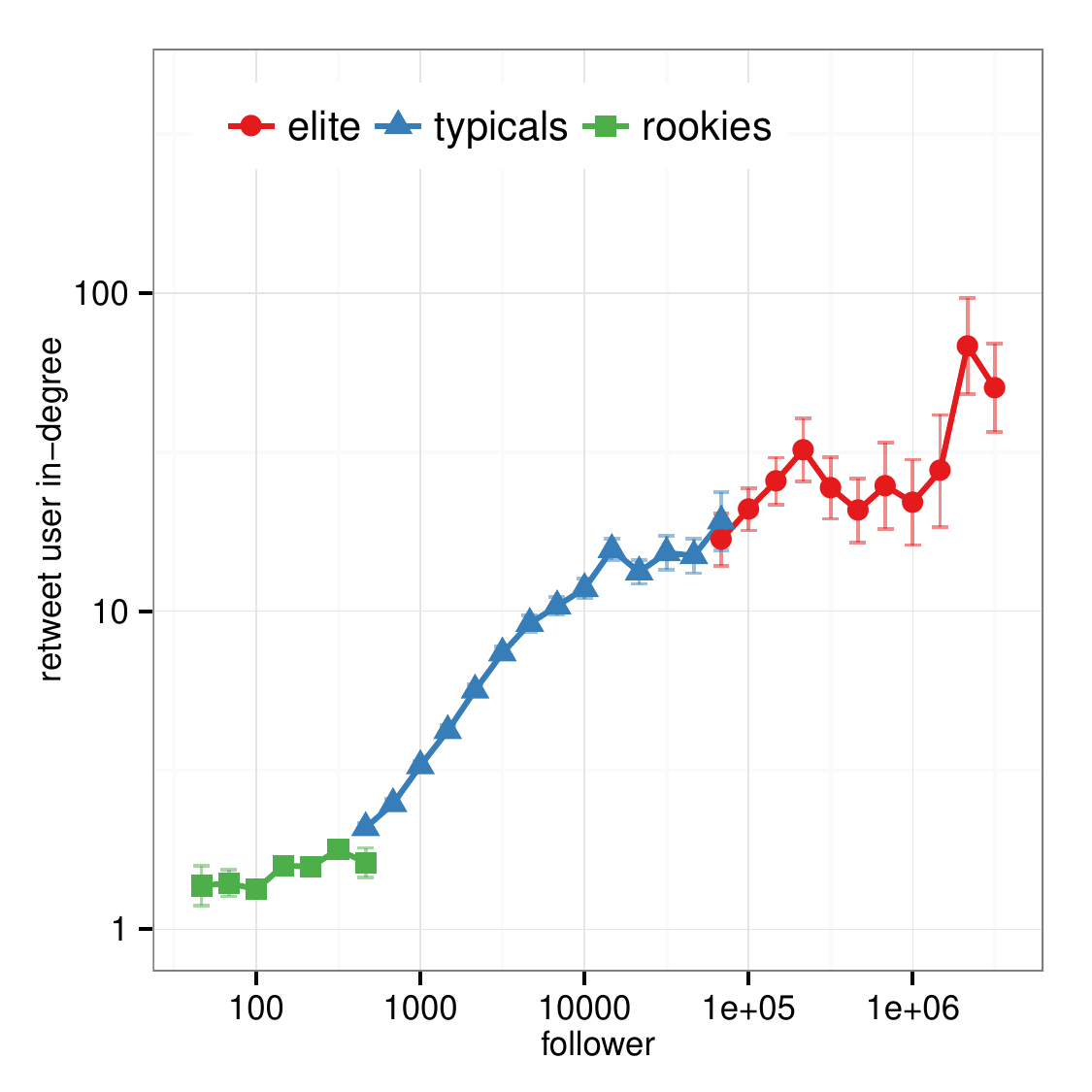} &
        \includegraphics[trim=.8cm 0cm .5cm 0cm, width=.24\textwidth]{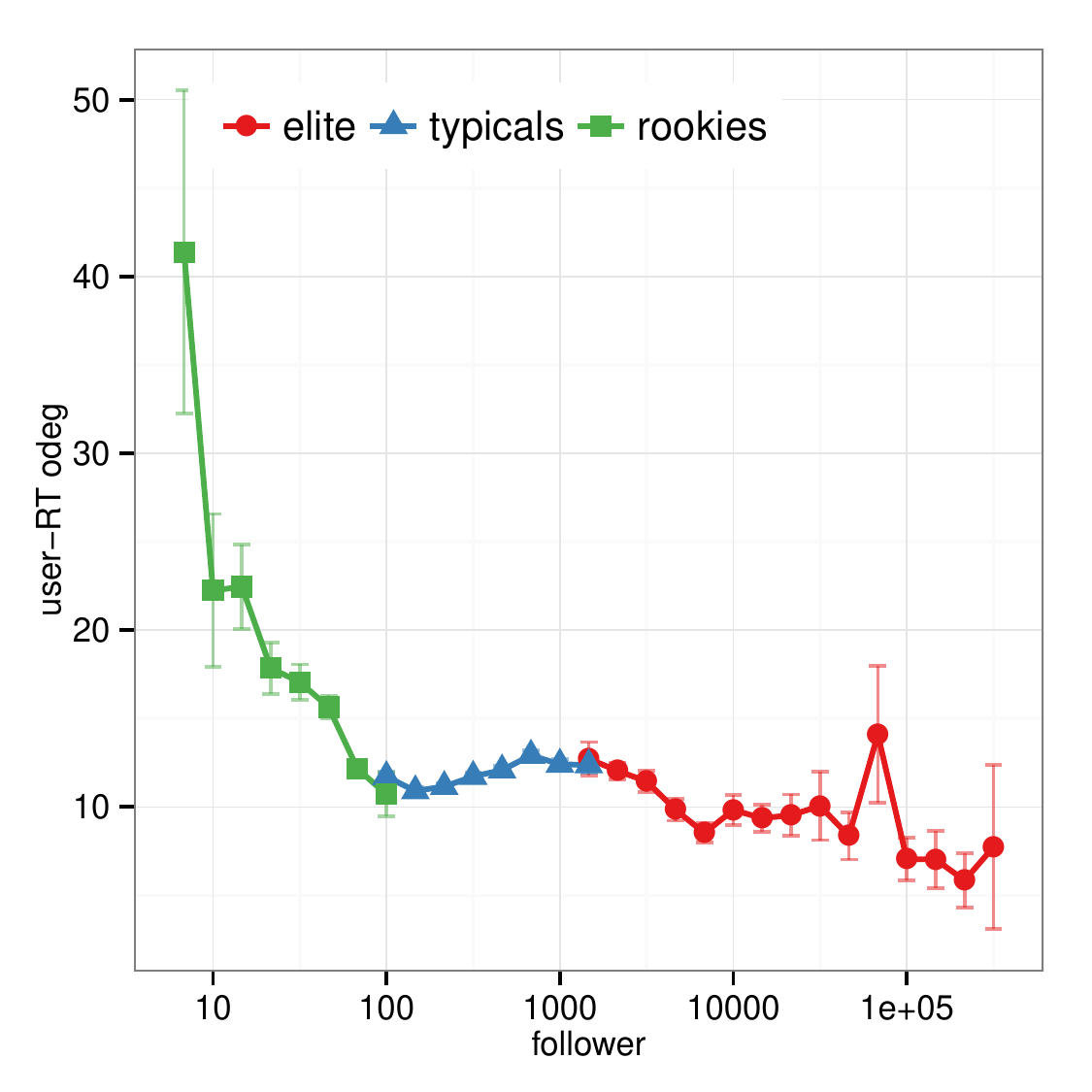} \\
(d) & (e) & (f)\\
        \includegraphics[trim=.8cm 0cm .5cm 0cm, width=.24\textwidth]{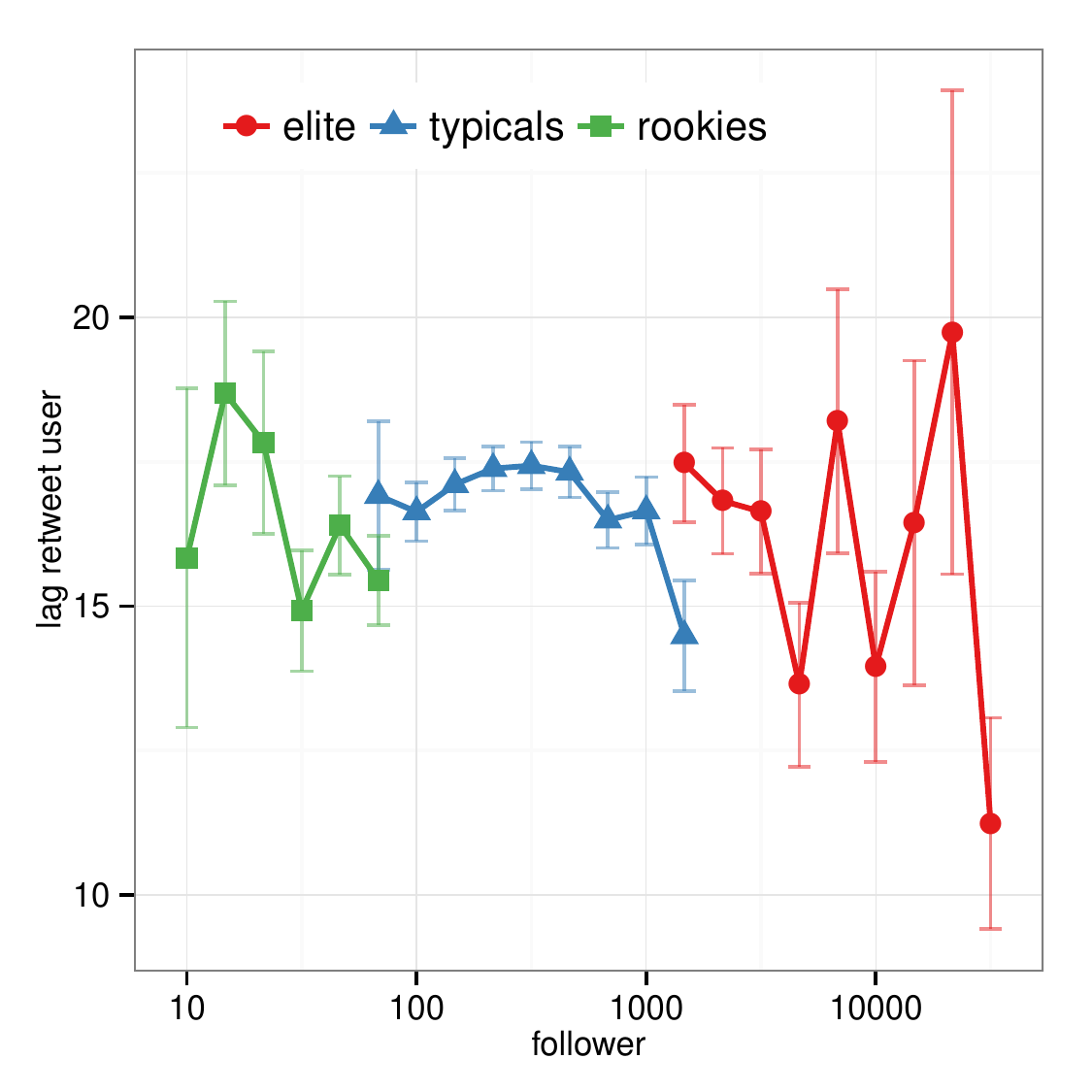} &
        \includegraphics[trim=.8cm 0cm .5cm 0cm, width=.24\textwidth]{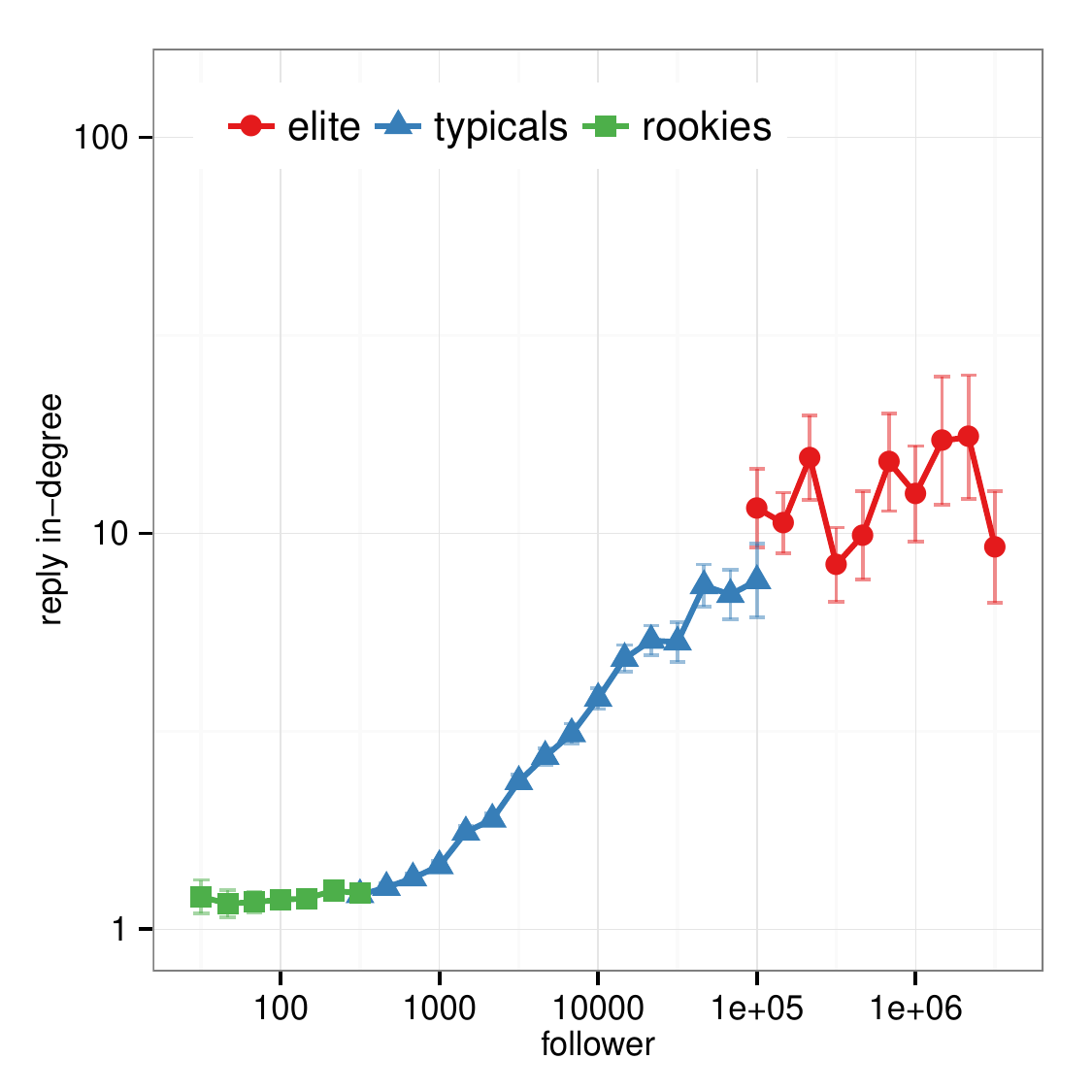} &
        \includegraphics[trim=.8cm 0cm .5cm 0cm, width=.24\textwidth]{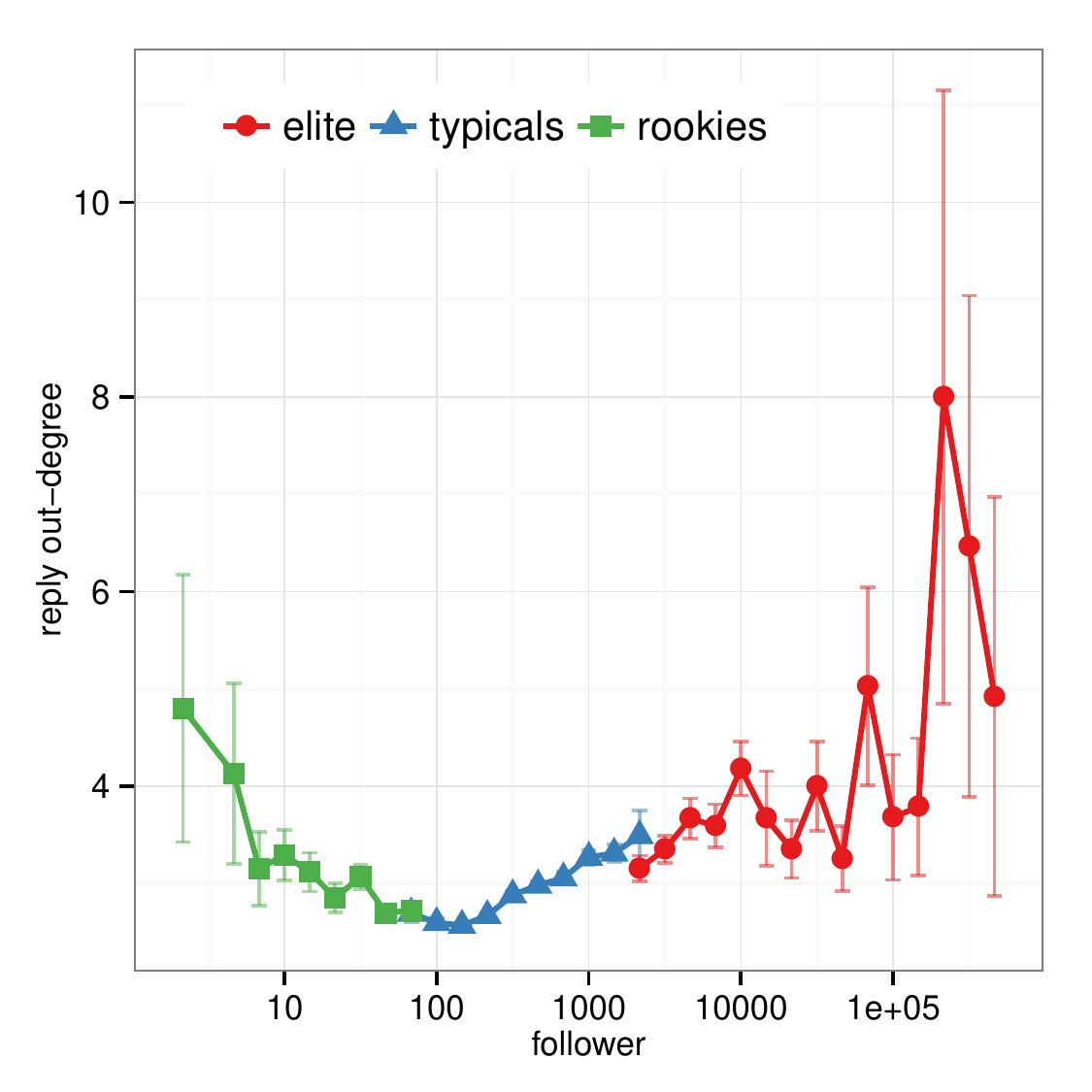} \\
\end{tabular}
\vspace{-1em}
    \caption{%
Responsiveness during debates. The $x$-axis plots the number of followers on a log-scale.
\label{fig:follower_segments}}\vspace{-0.5cm}
\end{figure*}

The prior sections examined the behavioral changes by aggregating all users together despite the significant differences in their activity. While activity became more centralized, it is unclear \textit{who} benefitted from this centralization: were users with larger audiences the focus of more activity? This section segments users into three classes based upon their audience size: ``elites'' are in the 90$th$ percentile for number of followers ($\ge 1805$), ``rookies'' are in the 10$th$ percentile for number of followers ($\le 88$), and ``typicals'' are the middle 80\%.

Based on this segmentation, Figure \ref{fig:follower_segments} plots the distributions for several of the activity types related to the concepts analyzed above. The ``tweet adoption index'' is the difference in average time to adopt retweet during the pre-debate baseline and the average time to retweet during the debate. High values for tweet adoption index indicate users adopted novel content in the debate more quickly than they did for the pre-debate baseline while low values indicate they adopted novel content at the same rate. For retweeting behavior, users with few followers retweeted much more rapidly during the debates than they did during the baseline and this behavior decreased as users' audience size grew (\ref{fig:follower_segments}a). 

Analogously, the ``user retweet adoption index'' is the difference in time to retweet a user during the pre-debate baseline and the time to retweet the same user in the debate. High values again indicate egos adopted alters' content more quickly during the debate than during normal time while low values indicate there is no difference in the time to adopt alters' content. Differences are less pronounced in the distribution of this index and audience size (\ref{fig:follower_segments}d). This suggests media events like the debate drive all users, regardless of audience size, to seek out and quickly retweet content from users they have previously retweeted.

Unsurprisingly, users with more followers are substantially more likely to have their content retweeted (\ref{fig:follower_segments}b) and to be mentioned by other users (\ref{fig:follower_segments}e) during the debates. Users with hundreds of thousands of followers retweeted less than 10 tweets during the debates while users with fewer than 10 followers retweeted more than 40 tweets during the debates (\ref{fig:follower_segments}c). While popular users retweeted less frequently than other types of users, they did engage in more interpersonal communication than users with fewer followers (\ref{fig:follower_segments}f).

Audience size has strong effects across the communication and information sharing activities we examined. These elite users retweet less frequently, adopt tweets more slowly, and reply much more often than other types of users during the peak hours of media events. This suggests elites' cultivation of large audiences is highly strategic: they only share some information with their audiences despite being the focus of substantial amounts of attention. At the other extreme, rookies are profligate and early adopters of novel content despite having little substantive impact on either their audiences.

\section{Discussion}
Previous work examining the dynamics of Twitter users' behavior generally assumed the underlying socio-technical system was stable and bursts of viral activity were functions of endogenous popularity dynamics related to the novelty or recency of the content. However, not all bursts of activity on social media are the same. Major media events like presidential debates or national political conventions are exogenous shocks that not only dominate the collective attention of Twitter users, but also significantly change the behavior of users and the system as a whole. While we found that information sharing behaviors like using hashtags or retweeting increased during these media events, interpersonal communication behaviors like mentioning and replies decreased in the aggregate. This lends support to our theory of media-event driven behavioral change that predicted users would enter more ``highly ordered'' states as conditions of shared attention and intense activity displaced the communication patterns of their normal social foci.

We examined whether these media events created ``rising tides'' that changed the behavior similarly across the system or if these events created ``rising stars'' that reinforced the attention and audience for already-elite users. While these media event-driven changes in activity are partially attributable to changes in behavior among the average user, they primarily reflect an increasing concentration of activity. References to hashtags, mentions of users, and retweets became significantly more centralized during media events without correspondingly large changes in the average behavior of users. Crucially, the beneficiaries of this newfound attention were not distributed throughout users with different numbers of followers, but concentrated among users with the largest audiences. Despite the potential for social media to increase our collective communicative capacity by having larger ``public squares'' with more diverse voices speaking, occasions for large-scale shared attention such as media events appear to replace existing social dynamics with increased collective attention to ``rising stars''.



The analyses we employed have several limitations that are opportunities for future work. Our data included only eight events across a relatively brief six-week period of time on topics related to politics, limiting the generalizability of these findings to other domains. Future work might explore whether similar patterns are found in other types of media events such as sports (\textit{e.g.}, Super Bowl) and awards ceremonies (\textit{e.g.}, Academy Awards) or across longer spans of time such as an entire political campaign. Despite the size of cohort of users whose behavior we analyzed and our intent to capture the behavior of politically-engaged users, the sampling strategy we employed potentially oversampled on users active during the debates. Alternative sampling strategies might uncover weaker or different social dynamics. A variety of more advanced metrics and features such as waiting times and assortative degree mixing could be used to analyze social dynamics in more detail. These analyses also involved quantitative analyses of aggregated behavioral traces that omitted analysis of the content of these tweets and the motivations of users that could be revealed by participant observation, interviews, and other qualitative methods.


\section{Conclusions}
This paper articulated a theory of \textit{media-event driven behavioral change} to explain for why some bursts of activity generate different patterns of behavior than other bursts. By considering not only changes in the overall level of activity, but changes in the structure of the networks of users, tweets, and hashtags, we identified the influence of several processes operating at micro- and macro- levels. Our findings demonstrate that changes in the aggregate levels of activity during media events are driven more by ``rising stars'' as elite users become the focus of collective attention rather than being driven by ``rising tides'' as users distribute their attention more broadly to new and diverse voices. These findings have implications for collective action under conditions of uncertainty and shared attention while suggesting new ways of mining information from social media to support collective sensemaking following major events. Social media like Twitter are not only sites for political communication among politicians and their supporters, they are increasingly becoming spaces for otherwise segmented audiences to come together in a third space to participate in consequential events.

\section{Acknowledgments}
Omitted for review.


\bibliographystyle{plain} \footnotesize
\bibliography{attention}

\end{document}